\documentclass[12pt,a4paper]{iopart}
\usepackage{epsfig}

\begin{document}

\title{Option pricing and perfect hedging on correlated stocks}

\author{Jaume Masoliver\footnote{Corresponding author. E-mail: {\tt jaume@ffn.ub.es}} and Josep Perell\'o}
\address{Departament de F\'{\i}sica Fonamental, Universitat de Barcelona, \newline Diagonal, 647, 08028-Barcelona, Spain}

\begin{abstract}

We develop a theory for option pricing with perfect hedging in an inefficient market model where the underlying price variations are autocorrelated over a time $\tau\geq 0$. This is accomplished by assuming that the underlying noise in the system is derived by an Ornstein-Uhlenbeck, rather than from a Wiener process. After obtaining an effective one-dimensional market model, we achieve a closed expression for the European call price within the Black-Scholes framework and find that our price is always lower than the Black-Scholes price. We obtain the same price if we start from a modified portfolio although now we get a different hedging strategy than that of Black-Scholes. We compare these strategies and study the sensitivity of the call price to several parameters where the correlation effects are also observed.

\end{abstract}



%\maketitle

\section{Introduction}

Fischer Black and Myron Scholes (1973) and Robert Merton (1973a) obtained
a fair option price assuming severe and strict theoretical conditions for
the market behavior. The requirements under which these
were developed include: (i) Absence of arbitrage opportunities, {\it
i.e.}, identical cashflows have identical values (Sharpe (1964); Cox and Ross
(1976)). (ii) Efficient market hypothesis, {\it i.e.}, the market
incorporates instantaneously any information concerning future market
evolution (Fama (1965)). (iii) Existence of a unique riskless strategy for
a portfolio in a complete market (Markowitz (1952)). Due to the random
character of stock market prices (Cootner (1964)), the implementation of
these conditions, especially condition (ii), indicates that speculative
prices are driven by white ({\it i.e.}, delta-correlated) random
processes. At this point, one has to choose between a Gaussian white
process (Black and Scholes (1973); Merton (1973a)) or a white jump
process. In this latter case and due to requirement (iii), the jump
lengths also have to be known and fixed (Merton (1976)). There are no
other choices for modelling market evolution if the above requirements and
ideal conditions are to be obeyed (Cox and Ross (1976)).

From these three assumptions, condition (ii) is perhaps the most
restrictive and, in fact, disagrees with empirical evidence since real
markets are not efficient, at least at short times (Grossman and Stiglitz
(1980); Fama (1991)). Indeed, market efficiency is closely related to
the assumption of totally uncorrelated price variations (white noise). But
white noise is only an idealization since, in practice, no actual random
process is completely white. For this reason, white processes are
convenient mathematical objects valid only when the observation time is
much larger than the autocorrelation time of the
process\footnote{Throughout this paper we will use the terms
``correlation" and ``autocorrelation" without distinction.}. And,
analogously, the efficient market hypothesis is again a convenient
assumption when the observation time is much larger than time spans in
which ``inefficiencies'' ({\it i.e.}, correlations, delays, etc....)
occur. 

Alternative models for describing empirical results of the market
evolution have been suggested (Mandelbrot (1963); Fama (1963)). In each
of these, an option price can be obtained only by relaxing some or even
all of the initial Black-Scholes (B-S) assumptions (Figlewski (1989);
Aurell \etal 
%, Baviera, Hammarlid, Serva, and Vulpiani
(2000)). Our main
purpose in this paper is to derive a nontrivial option price by relaxing
the efficient market hypothesis and allowing for a finite, non-zero,
correlation time of the underlying noise process. As a
model for the evolution of the market we choose the Ornstein-Uhlenbeck
(O-U) process (Uhlenbeck and Ornstein (1930)) for three reasons: (a) O-U
noise is still a Gaussian random process with an arbitrary correlation time
$\tau $ and it has the property that when $\tau=0$ the process becomes
Gaussian white noise, as in the original Black-Scholes option case. (b)
The O-U process is, by virtue of Doob's theorem, the only Gaussian random
process which is simultaneously Markovian and stationary (Doob (1942)). In
this sense the O-U process is the simplest generalization of Gaussian
white-noise. (c) As we will see later on, the variance of
random processes driven by O-U noise seems to agree with the evolution of
market variance, at least in some particular but relevant cases.

The Ornstein-Uhlenbeck process is not a newcomer in mathematical
finance. For instance, it has already been proposed as a
model for stochastic volatility\footnote{Hull and White (1987);
Scott (1987); Stein and Stein (1991); Heston (1993);
Ghysels \etal
%, Harvey and Renault
(1996); Heston and Nandi (2000).} (SV). Our case
here is rather different since, contrary to SV models, we only have
one source of noise. We therefore suggest the O-U process as the driving
noise for the underlying price dynamics when the volatility is
still a deterministic quantity (Dumas \etal 
%, Fleming, and Whaley 
(1998)).

The autocorrelation in the underlying driving noise is closely related to
the predictability of asset returns, of which there seems to be ample
evidence (Breen and Jagannathan (1989); Campbell and Hamao (1992)).
Indeed, if for some particular stock the 
price variations are correlated during some time $\tau$, then the price at
time $t_2$ will be related to the price at a previous time $t_1$ as long
as the time span $t_2-t_1$ is not too long compared to the correlation
time $\tau$. Hence correlation implies partial predictability.
Other approaches to option pricing with predictable asset returns are
based under the assumption the market is still driven by white noise and
predictability is induced by the drift (Lo and Wang (1995)). Since the B-S
formula is independent of the drift, these approaches apply B-S theory
with a conveniently modified volatility. Our approach here is rather
different because we assume the asset price variations driven by
correlated noise --which implies some degree of predictability.

Summarizing, our purpose is to study option pricing and hedging
in a more realistic framework that of white noise process presented by
Black and Scholes. Our model includes colored noise and the dependence of
the volatility on time. Both are empirically observed in real
markets (Bouchaud and Potters (2000)). Empirical characteristic time
scales are at least of the order of minutes and can affect option prices
particularly when the exercising date is near and speculative fluctuations
are more important. Presumably, this effect is negligible when
correlation times are shorter (much shorter than time to expiration). In
any case, it is interesting to know how, and by how much, the option price
and its properties are modified when correlations in the underlying noise
are significant.

The shortest way of getting the call price, and hence quantifying the
effect of correlations on prices is by martingale methods. Unfortunately,
this procedure does not guarantee that we obtain the fairest price since
arbitrage and hedging are not included in this approach. It is
therefore our main objective to generalize B-S theory not only to get a
new call price but, more importantly, to obtain a hedging strategy that
avoids risk and arbitrage opportunities.

From a technical point of view, we apply the B-S option pricing method
after projecting the two-dimensional O-U process onto a one-dimensional diffusion 
process with time varying volatility. As we will show, this projection allows 
us to maintain the conditions of a perfect hedging and the absence of 
arbitrage. Moreover, the price obtained using this way completely agrees with 
the price obtained using two alternative and different methods. One of them is
based on martingale theory, and the other one develops a new option pricing with
a modified portfolio containing secondary options instead of the underlying
stock.

The paper is divided into eight sections. In Section~\ref{sec:model} we present our
two-dimensional stochastic model for the underlying asset. In Section~\ref{sec:projected} we find the O-U projection onto the stock price correlated process. Section~\ref{sec:bs} concentrates on the B-S option price derivation with
the projected process, and Sections~\ref{sec:alt} and~\ref{sec:martingale} show the consistency of this derivation by using two alternative methods for obtaining the option price. The greeks and the new hedging are presented in Section~\ref{sec:hedging}. Conclusions are drawn in Section~\ref{sec:conc} and
technical details are left to the appendices.

\section{The asset model\label{sec:model}}

The standard assumption in option pricing theory is to assume that the
underlying price $S(t)$ can be modelled as a one-dimensional diffusion 
process:
\begin{equation}
\frac{dS(t)}{S(t)}=\mu dt+\sigma dW(t), 
\label{1a}
\end{equation}
where $W(t)$ is the Wiener process. In the original B-S theory both drift 
$\mu$ and volatility $\sigma$ are constants. Other
models take $\mu=\mu(t,S)$ and $\sigma=\sigma(t,S)$ as functions of time
and underlying price (Cox and Ross (1976); Bergman \etal
%, Grundy, and Wiener
(1987)). The parameter $\sigma$ is assumed to be a random quantity in the
SV models.

Notice that if the time evolution of the underlying price is governed by
Eq.~(\ref{1a}) then $S(t)$ is an uncorrelated random process in the sense
that its zero-mean return rate defined by $Z(t)=d\ln S/dt-\mu$ is
driven by white noise, {\it i.e.},
$E[Z(t_1)Z(t_2)]=\sigma^2\delta(t_1-t_2)$ where $\delta(t)$ is the Dirac
delta function\footnote{We recall that $\delta(x)$ is a generalized
function with the properties: $\delta(x)=0$ for $x\neq 0$ and
$\int_{-\infty}^{\infty}\delta(x)dx=1$ (Lighthill (1958); 
Stratonovich (1963)).}. 
Hence, the asset model immediately incorporates price return effects and 
meets the efficient market hypothesis. 

As a first step, we assume that the underlying price is not driven by the 
Wiener process $W(t)$ but by O-U noise $V(t)$. In other words, we say that
$S(t)$ obeys a singular two-dimensional diffusion 
\begin{equation}
\frac{dS(t)}{S(t)}=\mu dt+V(t)dt
\label{2a}
\end{equation}
\begin{equation}
dV(t)=-\frac{V(t)}{\tau}\ dt+
\frac{\sigma}{\tau}\ dW(t),
\label{3a}
\end{equation}
where $\tau\geq 0$ is the correlation time. More precisely, $V(t)$ is O-U
noise in the stationary regime, which is a Gaussian colored noise with
zero mean and correlation function:
\begin{equation}
E[V(t_1)V(t_2)]=\frac{\sigma^2}{2\tau}e^{-|t_1-t_2|/\tau}. 
\label{correlv}
\end{equation}

We call the process defined by Eqs.~(\ref{2a})-(\ref{3a}) singular
diffusion because, contrary to SV models, the Wiener driving noise $W(t)$
only appears in one of the equations, and this results in a singular
diffusion matrix (Gardiner (1985)). Observe that we now deal with autocorrelated
stock prices since the zero-mean return rate $Z(t)$ is colored noise,
{\it i.e.}, $E[Z(t_1)Z(t_2)]=(\sigma^2/2\tau) \exp[-|t_1-t_2|/\tau]$.
Note that when $\tau=0$ this correlation goes to $\sigma^2\delta(t_1-t_2)$ 
and we thus recover the one-dimensional diffusion discussed above.
Therefore the case of positive $\tau$ is a measure of the inefficiencies 
of the market. 

There is an alternative, and sometimes more convenient, way of writing the
above equations using the asset return $R(t)$ defined by 
\[
R(t)=\ln[S(t)/S_0],
\]
where $S_0=S(t_0)$ and $t_0$ is the time at which we start observing 
the process~(\ref{2a})-(\ref{3a}). Without loss of generality this time can be 
set equal to zero (see Appendix A). 
Instead of Eqs.~(\ref{2a})-(\ref{3a}), we may have 
\begin{equation}
\frac{dR(t)}{dt} =\mu+V(t)
\label{4a}
\end{equation}
\begin{equation}
\frac{dV(t)}{dt} = \frac{1}{\tau}\left[-V(t)+\sigma\xi(t)\right],
\label{4b}
\end{equation}
where $\xi(t)=dW(t)/dt$ is Gaussian
white noise defined as the derivative of the Wiener process. This process
exists in the sense of generalized random functions (Lighthill (1958);
Stratonovich (1963)). The combination of relations in Eqs.~(\ref{4a}) 
and~(\ref{4b}) leads to a second-order stochastic differential equation for 
$R(t)$ 
\begin{equation}
\tau\frac{d^2R(t)}{dt^2}+\frac{dR(t)}{dt}=\mu+\sigma\xi(t).
\label{6}
\end{equation}
From this equation, we clearly see that when $\tau=0$ we recover the
one-dimensional diffusion case~(\ref{1a})\footnote{In the opposite case
when $\tau=\infty$, Eq.~(\ref{3a}) shows that $dV(t)=0$. Thus $V(t)$ is a
constant, which we may equal to zero, and from Eq.~(\ref{2a}) we have
$S(t)=S_0e^{\mu t}$. Therefore, the underlying price evolves as a
riskless security. Later on we will recover this deterministic case (see,
for instance, Eq.~(\ref{deterministic})).} . We also observe that the O-U
process $V(t)$ is the random part of the return velocity, $dR/dt$, and we will
often refer to $V(t)$ as the ``velocity'' of the return process $R(t)$.

In Appendix A, we give explicit expressions for $V(t)$ and for the return 
$R(t)$. We prove there that $R(t)$ is a non-stationary process with the 
conditional mean value 
\begin{equation}
m(t,V_0)\equiv E[R(t)|V_0]=\mu t+\tau\left(1-e^{-t/\tau}\right)V_0,
\label{mr}
\end{equation}
where  $V(0)\equiv V_0$ is the initial
velocity. The conditional return variance,
\[
K_{11}(t)\equiv E[(R(t)-m(t,V_0))^2|V_0],
\]
is given by 
\begin{equation}
K_{11}(t)=
\sigma^2\left[t-2\tau\left(1-e^{-t/\tau}\right)+\frac{\tau}{2}
\left(1-e^{-2t/\tau}\right)\right].
\label{k}
\end{equation}
We also give in Appendix A explicit expressions for the joint
probability density function (pdf) $p(R,V,t)$, the marginal pdf's $p(R,t)$
and $p(V,t)$ of the second-order process $R(t)$, and the marginal pdf 
$p(S,t|S_0,t_0)$ of the underlying price $S(t)$. We also show that the
velocity $V(t)$ is, in the stationary regime, distributed 
according to the normal density:
\begin{equation}
p_{st}(V)=\frac 1{\sqrt{\pi \sigma ^2/\tau }}
e^{-\tau V^2/\sigma ^2}.
\label{pstat}
\end{equation}

Suppose now that the initial velocitiy $V_0$ is random with mean value 
$E[V_0]$ and variance ${\rm Var}[V_0]$. Thus, the return unconditional
mean and variance read
\begin{eqnarray*}
E[R(t)]&=&\mu t+\tau\left(1-e^{-t/\tau}\right)E[V_0],
\\
E[(R(t)-E(R(t))^2]&=&K_{11}(t)+\tau\left(1-e^{-t/\tau}\right){\rm Var}[V_0].
\end{eqnarray*}
If, in addition, we assume that the initial velocity $V_0$ is in the 
stationary regime then $E[V_0]=0$ and ${\rm Var}[V_0]= \sigma^2/2\tau$. 
In this case, the return unconditional mean value is 
\[
m(t)\equiv E[R(t)]=\mu t,
\]
and the return unconditional variance 
\[ 
\kappa(t) \equiv E[(R(t)-m(t))^2]
\]
reads ({\it cf.} Eq.~(\ref{k}))
\begin{equation}
\kappa(t)=
\sigma^2\left[t-\tau\left(1-e^{-t/\tau}\right)\right].
\label{varr}
\end{equation}

\begin{figure}
\begin{center}
\includegraphics[angle=-90,scale=0.55]{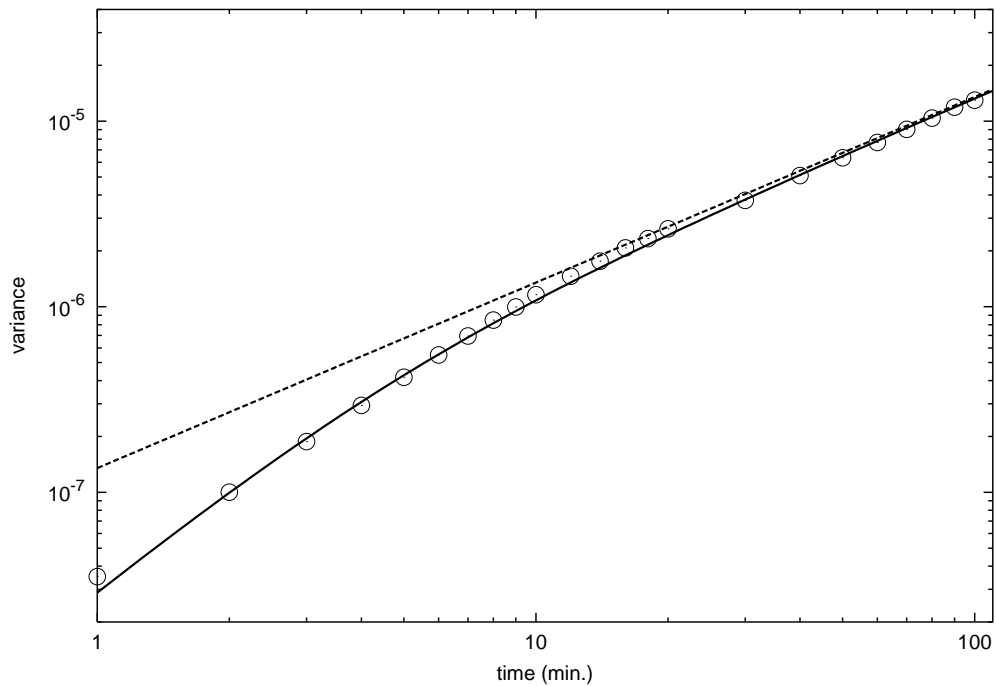}
\end{center}
\caption{The variance of the underlying asset as a function of time (in logarithmic scale). Circles correspond to empirical variance of S\&P 500 cash index from 1988 to 1996. Solid line represents the theoretical variance, Eq.~(7) with $\tau= 2$ minutes. The dashed line is the B-S variance $\sigma^2 t$. In both cases $\sigma=3.69\times 10^{-4} {\rm min}^{-1/2}$ which approximately corresponds to an annual volatility $\sigma=11\%$.}
\label{fig1}
\end{figure}

A consequence of Eq.~(\ref{varr}) is that, when $t\ll \tau$, the variance 
behaves as 
\begin{equation}
\kappa (t)\sim (\sigma ^2/2\tau )t^2,\qquad (t\ll \tau).
\label{superdif}
\end{equation}
Equation~(\ref{varr}) also shows a crossover to ordinary diffusion (B-S
case) when $t\gg\tau $: 
\begin{equation}
\kappa (t)\sim \sigma ^2t,\qquad (t\gg \tau ).
\label{dif}
\end{equation}
In Fig.~\ref{fig1}, we plot $\kappa (t)$ along with the empirical
variance from data of the S\&P 500 cash index during the period January
1988-December 1996\footnote{Tick by tick data on  S\&P 500 cash index has
been provided by The Futures Industry Institute (Washington, DC).}. The
dashed line represents results obtained by assuming normal-diffusion 
$\kappa (t)\propto t$. Observe that the empirical variance is very well fitted by our
theoretical variance $\kappa(t)$ for a correlation time $\tau =2$
minutes. Furthermore, the result of this correlation affects the empirical
volatility for around $100$ minutes. These times are probably too small to
affect call price to any quantifiable extent. However, the S\&P 500 is
one of the most liquid, and therefore most efficient, markets.
Consequently, the effect of correlations in any other less efficient
market might significantly influence option prices and hedging strategies, 
and this is the main motivation for this work.

\section{The projected process\label{sec:projected}}

One may argue that the O-U process~(\ref{2a})-(\ref{3a}) is an inadequate asset model since the share price $S(t)$ given by Eq.~(\ref{2a}) is a continuous random process with bounded variations. As Harrison \etal 
%, Pitbladdo and Schaefer 
(1984) showed, continuous processes with bounded variations allow arbitrage opportunities and this is an undesirable feature for obtaining a fair price. Thus, for instance, arbitrage would be possible within a portfolio containing bonds and stock whose strategy at time $t$ is buying (or selling) stock shares when $\mu+V(t)$ is greater (or lower) than the risk-free bond rate (Harrison \etal 
%, Pitbladdo and Schaefer 
(1984)). 

In our case, however, the problem is that in the practice {\it the return velocity $V(t)$ is nontradable and its evolution is ignored}. In other words, in real markets the observed asset dynamics does not show any trace of the velocity variable\footnote{Indeed, knowing $V(t)$ would imply knowing the value of the return $R(t)$ at two different times, since
\[
V(t)=\lim_{\epsilon\rightarrow 0^+}\frac{R(t)-R(t-\epsilon)}{\epsilon}-\mu.
\]
Obviously, this operation is not performed by traders who only manage portfolios at time $t$ based on prices at $t$ and not at any earlier time.}. This feature allows us to perform a projection of the two-dimensional diffusion process $[S(t),V(t)]$ onto a 
one-dimensional equivalent process $\bar{S}(t)$ independent of the velocity $V$. We will show in this section that the projected process $\bar{S}(t)$, which is equal to the actual price $S(t)$ in mean square sense, obeys the following one-dimensional SDE
\begin{equation}
\frac{d\bar{S}(t)}{\bar{S}(t)}=[\mu+\dot{\kappa}(T-t)/2] dt + 
\sqrt{\dot{\kappa}(T-t)} dW(t),
\label{sde}
\end{equation}
where $\kappa(t)$ is given by Eq.~(\ref{varr}), and the dot denotes time derivative. Therefore, the price given by Eq.~(\ref{sde}) is driven by a noise of unbounded variation, the Wiener process, and the Harrison \etal 
%, Pitbladdo and Schaefer 
(1984) results do not apply. In consequence, the O-U projected process is still a suitable starting point for option pricing since it does not permit arbitrage.

\subsection{Derivation of the one-dimensional SDE}

Note that the dynamics of the return $R(t)=\ln[S(t)/S_0]$ is given by the second-order SDE~(\ref{6}) which includes the stochastic evolution of the velocity $V(t)$. Let us now obtain a first-order SDE describing the price dynamics when velocity $V(t)$ has been eliminated.

The starting point of our derivation is the marginal conditional density
\newline $p(R,t|R_0,t_0;V_0)$. This density is given by Eq.~(\ref{pr0}) of Appendix A and when $t_0\neq0$ it reads
\begin{eqnarray}
p(R,t|R_0,t_0;V_0)=\frac{1}{\sqrt{2\pi K_{11}(t-t_0)}}
\exp\left\{-\frac{[R-R_0-m(t-t_0,V_0)]^2}{2K_{11}(t-t_0)}\right\},
\label{pr}
\nonumber \\
\end{eqnarray}
where $m(t,V_0)$ and $K_{11}(t)$ are given by Eqs.~(\ref{mr}) and~(\ref{k}). Note that \newline $p(R,t|R_0,t_0;V_0)$ is the solution of the following partial differential equation
\begin{equation}
\frac{\partial p}{\partial t_0} = 
-\left[\mu+V_0 e^{-(t-t_0)/\tau}\right] 
\frac{\partial p}{\partial R_0}
-\frac{\sigma^2}{2} \left[1-e^{-(t-t_0)/\tau}\right]^2 
\frac{\partial^2 p}{\partial R_0^2},
\label{bfpS}
\end{equation}
with the final condition $p(R,t|R_0,t;V_0)=\delta(R-R_0)$.
Observe that the Eq.~(\ref{bfpS}) is a backward Fokker-Planck equation whose 
drift, $\mu+V_0 \exp[-(t-t_0)/\tau]$, and diffusion coefficient,
$\frac{1}{2} \sigma^2 \left[1-\exp[-(t-t_0)/\tau]\right]^2$, are both 
functions of $t-t_0$. As is well-known, there exists a direct relation between
the Fokker-Planck equation and the SDE governing the process (Gardiner (1985)).
In our case, the corresponding SDE is
\begin{equation}
dR(t_0)=\left[\mu+V_0 e^{-(t-t_0)/\tau}\right] dt_0 
+\sigma \left[1-e^{-(t-t_0)/\tau}\right]dW(t_0),
\label{dRe}
\end{equation}
and its formal solution is
\begin{eqnarray}
R(t)= R(t_0)+\mu(t-t_0) &+& V_0\tau \left[1-e^{-(t-t_0)/\tau}\right]
\nonumber \\
&+& \sigma \int_{t_0}^{t} \left[1-e^{-(t-t_1)/\tau}\right]dW(t_1).
\label{Re}
\end{eqnarray}

\subsection{Equality of processes in mean square sense}

To avoid confusion, let $\bar{R}(t)$ be the solution of the first-order 
SDE~(\ref{dRe}), {\it i.e.}, $\bar{R}(t)$ is the projected process given by 
Eq.~(\ref{Re}). And let $R(t)$ be the solution of the second-order 
SDE~(\ref{6}) where the dynamics of the velocity is still taken into account. 
Thus, $R(t)$ is explicitly given by Eq.~(\ref{r2}) of Appendix A.

We will now prove that $\bar{R}(t)$ and $R(t)$ are equal in mean square 
sense. That is:
\begin{equation}
E\left[(R(t)-\bar{R}(t))^2\right]=0, \qquad \mbox{for any time } t.
\label{mean}
\end{equation}
In effect, from Eq.~(\ref{Re}) and assuming, without loss of 
generality, that $t_0=0$ and $\bar{R}(t_0)=0$ we have
\begin{equation}
\bar{R}(t)=\mu t+ V_0 \tau(1-e^{-t/\tau})
+\int_{0}^{t} \left[1-e^{-(t-t_1)/\tau}\right] \xi(t_1) dt_1,
\label{Re2}
\end{equation}
where $\xi(t_1)=dW(t_1)/dt_1$ is the Gaussian white noise. On the other hand, 
from Eq.~(\ref{r2}) we write
\begin{equation}
R(t)=\mu t+ V_0 \tau(1-e^{-t/\tau})
+\frac{\sigma}{\tau}\int_{0}^{t}dt'
\int_{0}^{t'}e^{-(t'-t^{\prime\prime})/\tau}
\xi(t^{\prime \prime})dt^{\prime\prime}.
\label{r2b}
\end{equation}
Therefore,
\begin{eqnarray*}
\fl E\Bigl[(R(t)-\bar{R}(t))^2\Bigr]= 2 K_{11}(t)
\\
\lo -\frac{2\sigma^2}{\tau}
\int_{0}^{t}dt'
\int_{0}^{t'}dt^{\prime\prime}e^{-(t'-t^{\prime\prime})/\tau} \int_{0}^{t}
dt_1 
\left[1-e^{-(t-t_1)/\tau}\right] E[\xi(t_1)\xi(t^{\prime \prime})],
\end{eqnarray*}
where $K_{11}(t)$ is given by Eq.~(\ref{k}). Taking into account that
\[
E[\xi(t')\xi(t)]=\delta(t'-t),
\]
we have
\[
E\left[(R(t)-\bar{R}(t))^2\right]= 2 K_{11}(t)-\frac{2\sigma^2}{\tau}
\int_{0}^{t}dt'\int_{0}^{t'}dt^{\prime\prime}e^{-(t'-t^{\prime\prime})/\tau} 
\left[1-e^{-(t-t^{\prime\prime})/\tau}\right].
\]
However, (see Eq.~(\ref{k}))
\[
\frac{\sigma^2}{\tau}
\int_{0}^{t}dt'\int_{0}^{t'}dt^{\prime\prime}e^{-(t'-t^{\prime\prime})/\tau} 
\left[1-e^{-(t-t^{\prime\prime})/\tau}\right]=K_{11}(t).
\]
Hence,
\[
E\left[(R(t)-\bar{R}(t))^2\right]=0,
\]
and $R(t)$ is equal to $\bar{R}(t)$ in mean square sense.

\subsection{The projected process when the initial velocity is in 
the stationary regime}

As we have mentioned, we are mainly interested in representing the asset
dynamics when the initial velocity $V_0$ is random and distributed according to the
stationary pdf~(\ref{pstat}). We have shown in Section~\ref{sec:model} that this basically implies the replacement of $K_{11}(t)$ by $\kappa(t)$. In such a case, the
SDE for $R(t)$ reads\footnote{Since $R(t)$ and $\bar{R}(t)$ are equal in 
mean square sense we will drop the bar on $\bar{R}$ as long as there is no 
confusion. Thus, we will use $R$ for the projected process as well.}
\[
dR(t)=\mu dt+\sqrt{\dot{\kappa}(T-t)}dW(t),
\]
where $\kappa(t)$ is given by Eq.~(\ref{varr}) and the dot denotes time 
derivative, that is
\begin{equation}
\dot{\kappa}(t)=\sigma^2\left(1-e^{-t/\tau}\right).
\label{dotvarr}
\end{equation}  
We need the It\^o lemma given in Appendix B for deriving the SDE for the 
stock $S$. Thus, according to Eq.~(\ref{itose1}), the effective dynamics 
for $S=S_0 e^R$ is   
\begin{equation}
\frac{dS(t)}{S(t)}=[\mu+\dot{\kappa}(T-t)/2] dt +
\sqrt{\dot{\kappa}(T-t)}dW(t).
\label{dSe}
\end{equation}
In this way, we have projected the two-dimensional O-U process $(S,V)$ onto a 
one-dimensional price process which is a Wiener process with time
varying drift and volatility. We also note that we need to specify the 
final condition of the process because the volatility $\sqrt{\dot{\kappa}}$ 
is a function of the time to maturity $T-t$, and this implies that the 
projected asset model depends on each particular contract.

\section{The option price on the projected process\label{sec:bs}} 
 
In this section we will present a generalization of the Black-Scholes
theory assuming that underlying price is driven by the O-U process. We
therefore eliminate the efficient market hypothesis but retain the other
two requirements of the original B-S theory: the absence of arbitrage and
the existence of a riskless strategy.

We invoke the standard theoretical restrictions --continuos trading without transaction costs and dividends-- and apply the original B-S method taking into account that the underlying asset is not driven by white noise but by colored noise modelled as an O-U
process.

The starting point of B-S option pricing is a portfolio which contains certain
amounts of shares, calls and bonds. In this context, B-S hedging is only able 
to remove the call risk that comes from stock fluctuations. Therefore, 
we need to start from the effective one-dimensional market dynamics given by 
Eq.~(\ref{dSe}) since otherwise we would not be able to remove risk 
fluctuations arising from $dW(t)$. These fluctuations are only explicitly 
given in the projected SDE for the stock (see Section~\ref{sec:2d} for a deeper discussion on this point).

\subsection{Black-Scholes option pricing with the equivalent one-dimensional SDE}

As we have proved in Section~\ref{sec:projected}, there exists an effective one-dimensional diffusion which describes the O-U process~(\ref{2a})-(\ref{3a}). Assuming that the effective one-dimensional price dynamics is given by Eq.~(\ref{dSe}), it is quite straightforward to 
derive the European call option price within the original B-S method. Following Merton (1973b) we define a portfolio compounded by a certain amount $\Delta $ of shares at
price $S$, a quantity of bonds $\Phi$, and a number $\Psi $ of calls with price $C$, maturity time $T$ and strike price $K$. We assume that short-selling is allowed and thus the value $P$ of the portfolio is written
\begin{equation}
P=\Psi C-\Delta S-\Phi B,
\label{P-B-S}
\end{equation}
where the bond price $B$ evolves according to the risk-free interest rate
ratio $r$. That is
\begin{equation}
dB=r Bdt.
\label{bond}
\end{equation}
The portfolio is required to obey the net-zero investment hypothesis,
which means $P=0$ for any time $t$ (Merton (1973b)). Hence, 
\begin{equation}
C=\delta S+\phi B,
\label{C-B-S}
\end{equation}
where $\delta=\Delta/\Psi$ and $\phi=\Phi/\Psi$ are, respectively,
the number of shares per call and the number of bonds per call. Due to the
nonanticipating character of $\delta$ and $\phi$ we have (Bj\"{o}rk (1998))
\begin{equation}
dC=\delta dS+\phi dB.
\label{nonant}
\end{equation}
On the other hand, assuming that the market dynamics is described by 
Eq.~(\ref{dSe}), the differential of the call also reads
\begin{equation}
dC(S,t)= C_t dt +C_S dS+\frac{1}{2} \dot{\kappa}(T-t)S^2C_{SS} dt,
\label{sde2}
\end{equation}
where we have used the It\^o lemma as expressed by Eq.~(\ref{itose}) of the
Appendix B. From  Eqs.~(\ref{nonant})-(\ref{sde2}) and~(\ref{C-B-S}) we get
\[
\left[C_t +\frac{1}{2} \dot{\kappa}(T-t)S^2C_{SS}+r\delta S -rC\right] dt
=[\delta -C_S]dS.
\]
Now the B-S delta hedging, $\delta=C_S$,
removes any random uncertainty in the option price.
The partial differential equation for $C(S,t)$ then reads
\begin{equation}
C_t=rC-rSC_S-\frac{1}{2}\dot{\kappa}(T-t)S^2C_{SS}.
\label{sde3}
\end{equation}
We note that the delta hedging is able to remove risk because
we have projected the two-dimensional SDE~(\ref{2a})-(\ref{3a}) onto the 
one-dimensional process. In this way, we directly relate the differential of 
the stock $dS(t)$ to the random fluctuations of the Wiener process $dW(t)$ 
(see Eq.~(\ref{dSe})). Without this projection, the B-S hedging is useless 
and the random fluctuations persist in the B-S portfolio. We will further 
discuss this situation in Section~\ref{sec:2d}.

\subsection{The price of the European call}

For the European call, Eq.~(\ref{sde3}) has to be solved with the following ``final condition" at maturity time $T$
\begin{equation}
C(S,T)=\max[S(T)-K,0], 
\label{finalC}
\end{equation}
where $S(T)$ is the underlying price at maturity and $K$ is the strike
price. The solution to Eq.~(\ref{sde3}) subject to Eq.~(\ref{finalC}) is
a type of solution perfectly known in the literature (see, for instance, 
Hull (2000)). Thus, our final price is
\begin{equation}
C_{OU}(S,t)=S\ N(d^{OU}_1)-Ke^{-r (T-t)}\ N(d^{OU}_2),
\label{call4} 
\end{equation}
where
\[
N(z)=(1/\sqrt{2\pi})\int_{-\infty}^{z}e^{-x^2/2}dx
\]
is the probability integral, and
\begin{equation}
d^{OU}_1=\frac{\ln(S/K)+r (T-t)+\kappa(T-t)/2}{\sqrt{{\kappa}(T-t)}}, 
\label{d1}
\end{equation}
\begin{equation}
d^{OU}_2=d^{OU}_1-\sqrt{{\kappa}(T-t)},
\label{d2} 
\end{equation}
with $\kappa (t)$ given by Eq.~(\ref{varr}).

Equation~(\ref{call4}) constitutes the key result of the paper. Note that,
when $\tau=0$, the variance becomes $\kappa(t)=\sigma^2 t$ and the 
price in Eq.~(\ref{call4}) reduces to the Black-Scholes price: 
\begin{equation}
C_{BS}(S,t)=S\ N(d_1^{BS})-Ke^{-r (T-t)}\ N(d_2^{BS}),
\label{call4B-S} 
\end{equation}
where $d_{1,2}^{BS}$ have the form of Eqs.~(\ref{d1})-(\ref{d2}) with $\kappa(T-t)$
replaced by $\sigma^2 (T-t)$. Therefore, the O-U price in
Eq.~(\ref{call4}) has the same functional form as B-S price in
Eq.~(\ref{call4B-S}) when $\sigma^2t$ is replaced by $\kappa(t)$.

In the opposite case, $\tau=\infty$, where there is no random noise but a
deterministic and constant driving force (in our case it is zero),
Eq.~(\ref{call4}) reduces to the deterministic price 
\begin{equation}
C_{d}(S,t)=\max\left[S-Ke^{-r (T-t)},0\right].
\label{deterministic} 
\end{equation}

\begin{figure}
\begin{center}
\includegraphics[angle=-90,scale=0.55]{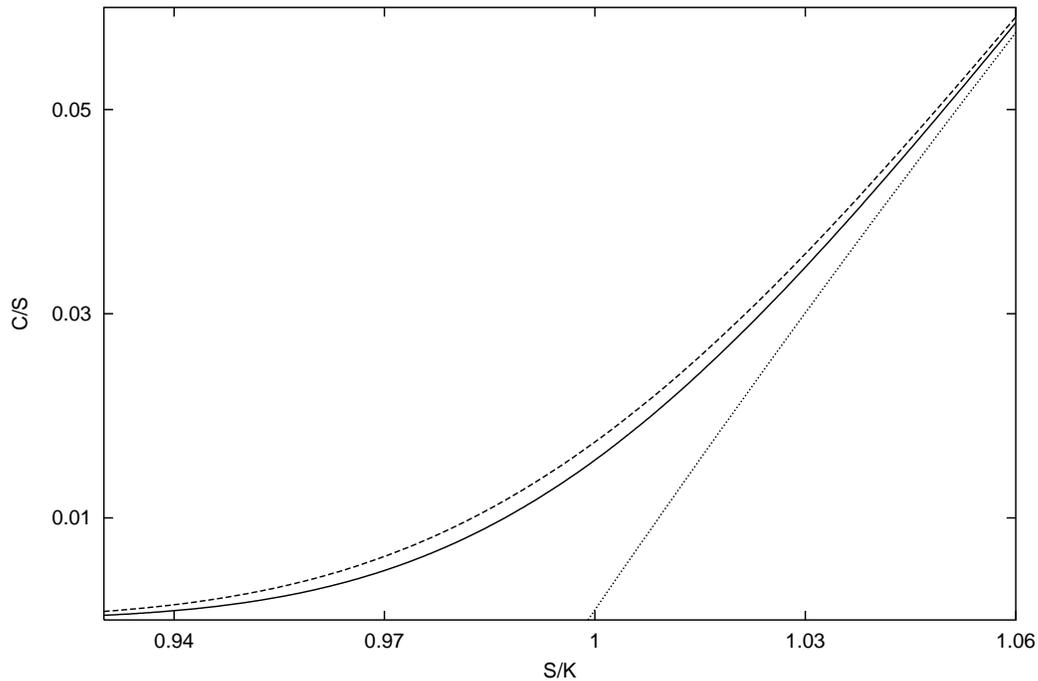}
\end{center}
\caption{Relative call price $C/S$ as a function of $S/K$ for a given time to expiration $T-t=5$ days. The solid line represents the O-U call price with $\tau=1$ day and the dashed line is the B-S price. The dotted line is the deterministic price. In this figure the annual risk-free interest rate $r=5 \%$, and the annual volatility $\sigma=30 \%$.}
\label{fig2}
\end{figure}

We will now prove that $C_{OU}$ is an intermediate price between B-S price and
the deterministic price (see Fig.~\ref{fig2})
\begin{equation}
C_{d}(S,t)\leq C_{OU}(S,t)\leq C_{BS}(S,t),
\label{bounds} 
\end{equation}
for all $S$ and $0\leq t\leq T$. In order to prove this it suffices to
show that $C_{OU}$ is a monotone decreasing function of the correlation
time $\tau$, since in such a case 
\[
C_{OU}(\tau=\infty)\leq C_{OU}(\tau)\leq C_{OU}(\tau=0).
\]
However, $C_{OU}(\tau=\infty)=C_d$ and $C_{OU}(\tau=0)=C_{BS}$, which
leads to Eq.~(\ref{bounds}). Let us thus show
that $C_{OU}$ is a decreasing function of $\tau$ for $0\leq t\leq T$ and
all $S$. Define a function $\alpha$ as the derivative  
\begin{equation}
\alpha=\frac{\partial C_{OU}}{\partial\tau}.
\label{nu}
\end{equation}
Since the $\tau$ dependence in $C_{OU}$ is a consequence of the variance
$\kappa(t,\tau)$, we have
\[
\alpha=\frac{\sigma}{2\kappa(T-t,\tau)}
\frac{\partial\kappa(T-t,\tau)}{\partial\tau}
{\cal V}_{OU},
\]
where ${\cal V}_{OU}=\partial C_{OU}/\partial\sigma$ (see Section~\ref{sec:hedging}). But 
\[
\frac{\partial\kappa(T-t,\tau)}{\partial\tau}=
-\sigma^2\left[1-(1+(T-t)/\tau)e^{-(T-t)/\tau}\right]\leq 0,
\]
for $0\leq t\leq T$ which is seen to be non positive. From
Eq.~(\ref{vega}) below we see that 
${\cal V}_{OU}\geq 0$ for all $S$ and $0\leq t\leq T$. Hence,
$\alpha\leq 0$ which proves Eq.~(\ref{bounds}). In Fig.~\ref{3a} we plot 
the option price $C$ as a function of the correlation time $\tau$ and 
for three different values of the moneyness $S/K$. This figure clearly 
shows that $C$ is a monotone decreasing function of $\tau$. 

\begin{figure}
\begin{center}
\includegraphics[angle=-90,scale=0.55]{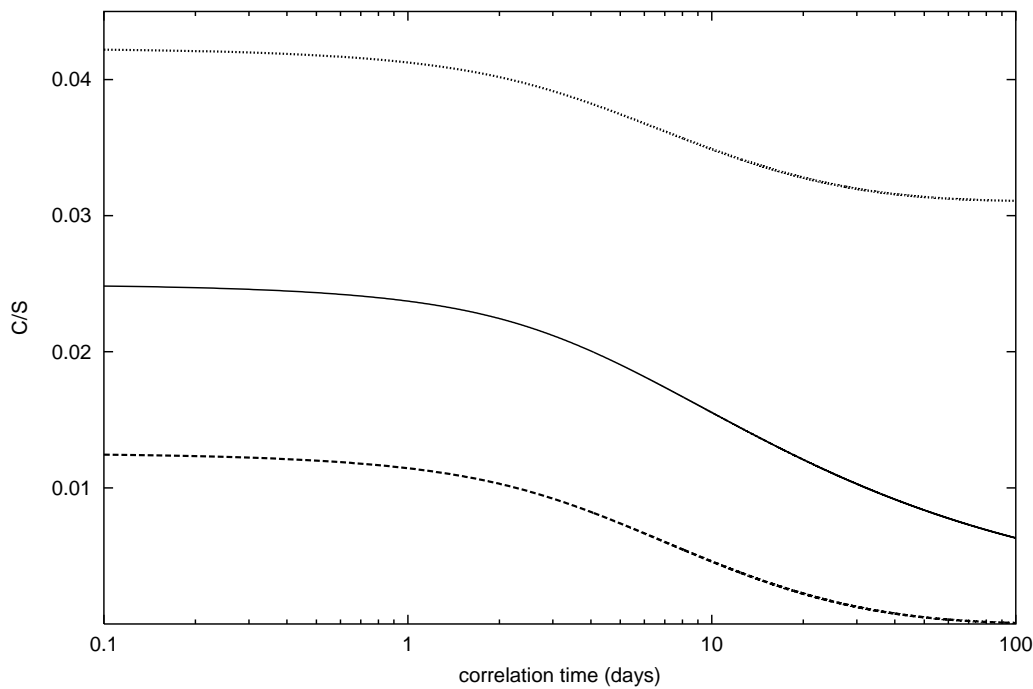}
\end{center}
\caption{Relative call price $C/S$ as a function of $\tau$ for a given time to expiration $T-t=10$ days. The solid line represents the call price with $S/K=1$ (ATM case). The dotted line is the call price when $S/K=1.03$ (ITM case). The dashed line represents an OTM case when $S/K=0.97$. We clearly see that $C$ is a monotone decreasing function of $\tau$ having its maximum value when $\tau=0$ (B-S case) and its minimum when $\tau\rightarrow\infty$ (deterministic price). The annual risk-free interest rate and the annual volatility are as in Fig. 2.}
\label{fig3a}
\end{figure}

\begin{figure}
\begin{center}
\includegraphics[angle=-90,scale=0.55]{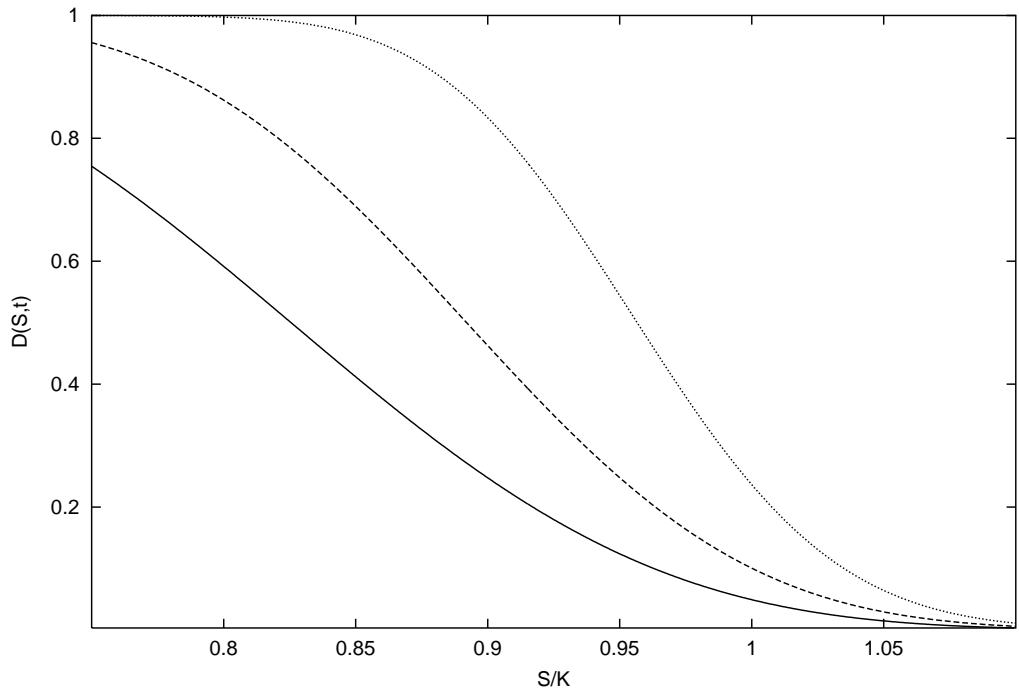}
\end{center}
\caption{$D(S,t)$ is plotted as a function of $S/K$ for $T-t=10$ days and $\tau=1$ day (solid line), $\tau=2$ days (dashed line) and $\tau=5$ days (dotted line). Other parameters used to generate the figure are $r=5 \%$ per annum and $\sigma=30 \%$ per annum.}
\label{fig3}
\end{figure}

Therefore, {\it the assumption of uncorrelated underlying
assets (B-S case) overprices any call option}. This confirms
the intuition understanding that correlation implies more predictability
and therefore less risk and, finally, a lower price for the option. In
fact, we can easily quantify this overprice by evaluating the relative
difference
\[
D=(C_{BS}-C_{OU})/C_{BS}.
\]
Figure~\ref{fig3} shows the ratio $D(S,t)$, for a fixed time to
expiration, plotted as a function of the moneyness, $S/K$, and for different
values of correlation time $\tau$. We see there that the ratio $D$ is very
sensitive to whether the call is in the money (ITM), out of the money
(OTM) or at the money (ATM). The biggest difference between prices occurs
in the case of OTM options. This is true because when $S/K<1$, both
$C_{BS}$ and $C_{OU}$ are small but $C_{BS}\gg C_{OU}$ (see
Fig.~\ref{fig2}). Depending on the value of correlation time $\tau$ this
implies that $D$ is approximately equal to 1.

\begin{figure}
\begin{center}
\includegraphics[angle=-90,scale=0.55]{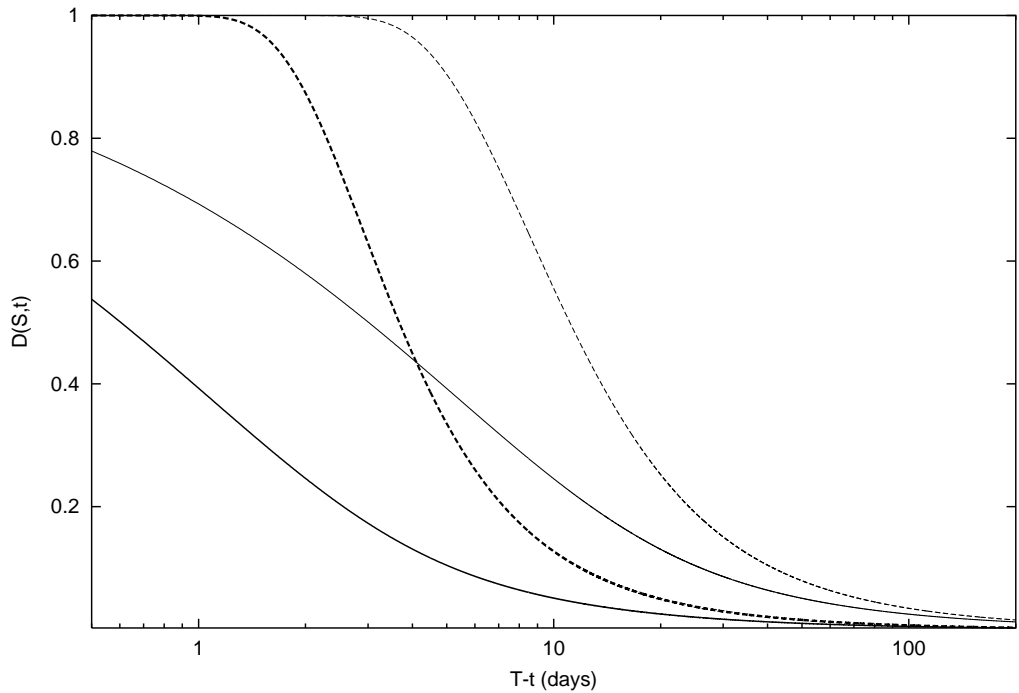}
\end{center}
\caption{$D(S,t)$ is plotted as a function of $T-t$ (in logarithmic scale) for fixed values of moneyness. The solid lines represent ATM options, the thick line corresponds to $\tau=1$ day and the thin line corresponds to $\tau=5$ days. The dashed lines represent an OTM option with $S/K=0.95$, the thick line corresponds to $\tau=1$ day and the thin line corresponds to $\tau=5$ days ($r$ and $\sigma$ as in Fig. 2).}
\label{fig4}
\end{figure}

\begin{figure}
\begin{center}
\includegraphics[angle=-90,scale=0.55]{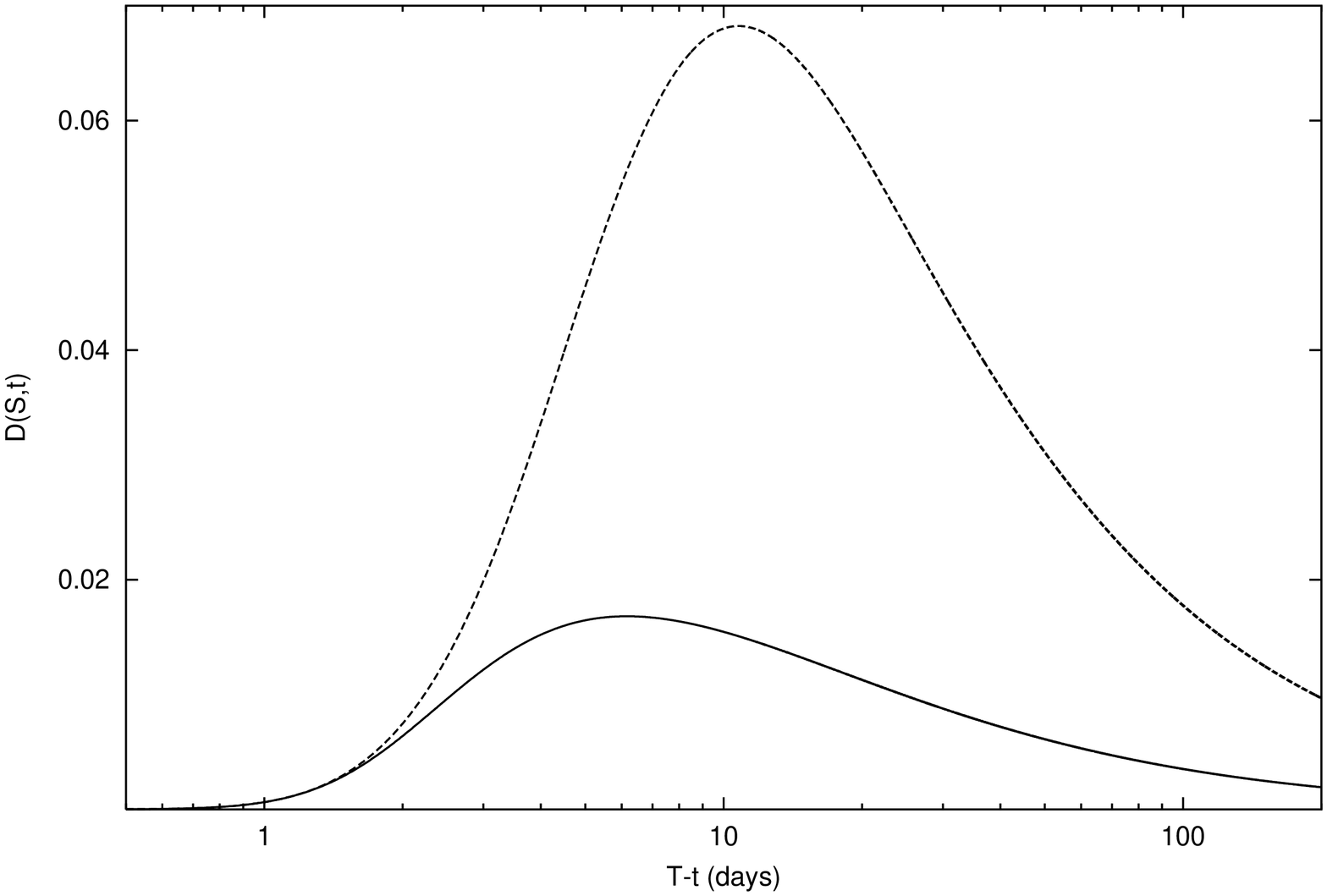}
\end{center}
\caption{$D(S,t)$ is plotted as a function of the expiration time $T-t$ (in logarithmic scale) for an ITM option with $S/K=1.05$. The solid line corresponds to $\tau=1$ day and the dashed line to $\tau=5$ days ($r$ and $\sigma$ as in Fig. 2).}
\label{fig5}
\end{figure}

Another interesting point is the behavior of $D$ as a function of the
expiration time $T-t$. In this case, $D$ behaves quite differently
depending on whether the call is in, out, or at the money. This behavior
is evident in Figs.~\ref{fig4} and~\ref{fig5}.
Figure~\ref{fig4} shows $D(S,t)$ as a function of expiration time $T-t$
for an OTM option ($S/K=0.95$) and the ATM option ($S/K=1.00$) and for two
different values (1 and 5 days) of the correlation time. Note that B-S
notably overprices the option, particularly in the OTM case. In 
Fig.~\ref{fig5} we show plots of $D(S,t)$ as a function of $t$ for an ITM
option ($S/K=1.05$). This exhibits completely different
behavior since the B-S overprice is considerably less (no
more than 7\%). Moreover, contrary to the ATM and OTM cases, the
relative difference $D(S,t)$ is a non monotone function of $T-t$,
having a maximum value around one or two weeks before maturity.
Although perhaps the most striking and interesting feature is {\it the
persistence of the B-S overprice far from maturity} regardless the value
of the correlation time. This is clearly
shown in Table~\ref{tab1} where we quantify the ratio $D$ in percentages 
for different values of moneyness, time to expiration and correlation
time.

\begin{table}
\caption{Relative call price differences in percentages. Values of $D\times 100$, where $D=(C_{BS}-C_{OU})/C_{BS}$. $T-t$ is the expiration time in days. Correlation times $\tau$ are 1, 2, and 5 days. The rest of columns are divided in three blocks corresponding to a different values of the moneyness $S/K$. From left to right blocks represent the OTM, ATM, and ITM cases. Notice the importance and the persistence far from maturity of the relative differences in price ($r$ and $\sigma$ as in Fig. 2).}\label{tab1}
\begin{tabular*}{\textwidth}{@{}l*{15}{@{\extracolsep{0pt plus12pt}}l}}
\br
$T-t$ & & \multicolumn{3}{c}{$S/K=0.95$} & &
\multicolumn{3}{c}{$S/K=1.00$} & &
\multicolumn{3}{c}{$S/K=1.05$} \\
& & $\tau=$1 & 2 & 5 & &
$\tau=$1 & 2 & 5 & &
$\tau=$1 & 2 & 5 \\
\mr
  1 & & 99.9 & 100 & 100 & & 39.3 & 53.8 & 69.3 & & 0.1 & 0.1 & 0.1 \\
  2 & & 87.4 & 98.7 & 100 & & 24.6 & 39.3 & 58.0 & & 0.6 & 0.7 & 0.7 \\
  3 & & 62.9 & 88.3 & 99.5 & & 17.3 & 30.5 & 50.1 & & 1.2 & 1.7 & 2.0 \\
  4 & & 45.0 & 73.1 & 96.5 & & 13.1 & 24.6 & 44.1 & & 1.5 & 2.5 & 3.4 \\
  5 & & 33.5 & 59.4 & 90.4 & & 10.5 & 20.4 & 39.2 & & 1.6 & 2.9 & 4.6 \\
  6 & & 26.1 & 48.6 & 82.8 & & 8.7 & 17.3 & 35.3 & & 1.7 & 3.1 & 5.5 \\
  7 & & 21.0 & 40.4 & 75.1 & & 7.4 & 14.9 & 31.9 & & 1.7 & 3.2 & 6.1 \\
  8 & & 17.4 & 34.1 & 67.9 & & 6.4 & 13.1 & 29.1 & & 1.6 & 3.2 & 6.5 \\
  9 & & 14.7 & 29.2 & 61.3 & & 5.7 & 11.6 & 26.7 & & 1.6 & 3.2 & 6.7 \\
  \\ 
  10 & & 12.7 & 25.4 & 55.6 & & 5.1 & 10.4 & 24.5 & & 1.5 & 3.1 & 6.8 \\
  20 & & 5.0 & 10.0 & 25.2 & & 2.5 & 5.1 & 13.1 & & 1.1 & 2.3 & 5.7 \\    
  30 & & 2.9 & 5.9 & 15.1 & & 1.7 & 3.4 & 8.6 & & 0.9 & 1.8 & 4.5 \\ 
  40 & & 2.0 & 4.1 & 10.5 & & 1.2 & 2.5 & 6.4 & & 0.7 & 1.4 & 3.7 \\ 
  50 & & 1.6 & 3.1 & 7.9 & & 1.0 & 2.0 & 5.1 & & 0.6 & 1.2 & 3.1 \\
  \\
  100 & & 0.7 & 1.4 & 3.4 & & 0.5 & 1.0 & 2.5 & & 0.4 & 0.7 & 1.8 \\
  150 & & 0.4 & 0.8 & 2.2 & & 0.3 & 0.7 & 1.6 & & 0.2 & 0.5 & 1.3 \\ 
  200 & & 0.3 & 0.6 & 1.5 & & 0.2 & 0.5 & 1.2 & & 0.2 & 0.4 & 1.0 \\  
  250 & & 0.2 & 0.5 & 1.2 & & 0.2 & 0.4 & 1.0 & & 0.2 & 0.3 & 0.8 \\ 
\br
\end{tabular*}
\end{table}

\section{An alternative derivation of the call price\label{sec:alt}}

In this section and the next, we present two different and alternative 
derivations of the final call price $C_{OU}$. The first of these 
derivations is based on an extension of the B-S theory but now starting from 
the two-dimensional diffusion~(\ref{2a})-(\ref{3a}) and with a different 
portfolio than the usual one. A second derivation, briefly outlined in the next
section, uses the equivalent martingale measure method. 
Both derivations arrive at the price formula~(\ref{call4}), 
thus showing the consistency of the pricing methods.

We will first apply the original B-S method starting from the
two-dimensional O-U process~(\ref{2a})-(\ref{3a}) instead of the equivalent 
process~(\ref{dSe}). Unfortunately, this procedure yields a trivial expression
for the price of the option (see below) and is therefore useless. To avoid this
difficulty we will define a different portfolio which is the first step towards
the generalization of both B-S equation and formula.

\subsection{The Black-Scholes method for the two-dimensional O-U process\label{sec:2d}}

We assume that market prices are driven by an O-U process as shown in 
Eqs.~(\ref{2a})-(\ref{3a}) and that the portfolio is given by 
Eq.~(\ref{P-B-S}). That is, $C=\delta S+\phi B$ and
\[
dC=\delta dS+\phi dB.
\]
Let us now apply the original B-S method starting from the two-dimensional O-U 
process~(\ref{2a})-(\ref{3a}) instead of the equivalent process~(\ref{dSe}).
Using the It\^o lemma for a singular two-dimensional diffusion 
(see Appendix B), 
\begin{equation}
dC(S,V,t)=C_SdS+C_VdV+C_tdt+\frac{\sigma ^2}{2\tau ^2}C_{VV}dt,
\label{ito}
\end{equation}
and taking Eqs.~(\ref{bond}) and~(\ref{nonant}) into account, we write 
\[
\left[ C_t+\frac{\sigma ^2}{2\tau }C_{VV}-
r (C-S\delta)\right]dt+(C_S-\delta )dS+C_VdV=0.
\label{B-S}
\]
Now the assumption of delta hedging $\delta=C_S$, turns this equation into 
\begin{equation}
\left[ C_t-r (C-SC_S)+\frac{\sigma ^2}{2\tau }C_{VV}\right]
dt+C_VdV=0.
\label{B-S1}
\end{equation}
Equation~(\ref{B-S1}) is still random due to the term with $dV$ representing
velocity fluctuations (see Eq.~(\ref{3a})). In consequence, Black-Scholes delta 
hedging is incomplete since it is not able to remove risk. 
In this situation, the only way to derive a risk-free partial differential 
equation for the call price is to assume that the call is independent of velocity.
Then, $C_V=0$ and Eq.~(\ref{B-S1}) yields 
\begin{equation}
C_t+r SC_S-r C=0. 
\label{deteq}
\end{equation}
According to the final condition for the European call,  
$C(S,T)=\max [S(T)-K,0]$, the call price is
$C(S,t)=\max\left[S-Ke^{-r (T-t) },0\right]$. Note that this is a
useless expression because it gives a price for the option as if the
underlying asset would have evolved deterministically like the risk-free
bond without pricing the random evolution of the stock. In fact, there is
no hint of randomness, measured by the volatility $\sigma$, in
Eq.~(\ref{deteq}).

\setcounter{footnote}{1}

The main reason for the failure of B-S theory is the
inappropriateness of B-S hedging for two-dimensional processes such as
O-U price process~(\ref{2a})-(\ref{3a})\footnote{A similar situation appears
in the stochastic volatility models (Scott (1987)).}. Indeed, delta hedging
presumably diversifies away the risk associated with the differential of
asset price $dS(t)$ given by Eq.~(\ref{2a}). Nevertheless, what we have
to hedge is the risk associated with $dV(t)$ given by Eq.~(\ref{3a}), which 
contains the only source of randomness: the differential of the Wiener process
$dW(t)$. All of this clearly shows the uselessness of the B-S delta hedging for 
the two-dimensional O-U process. Note that we must relate in a direct way 
the differential $dS(t)$ with the random differential $dW(t)$, 
otherwise we will not be able to remove risk. This is indeed the case of 
the projected process~(\ref{dSe}) which leads to the European call price, 
Eq.~(\ref{call4}). However, if we do not want to project the 
process and maintain the two-dimensional formulation~(\ref{2a})-(\ref{3a}) 
we have to evaluate the option price from a different portfolio.
We will do it next by defining a new portfolio which will allow us to preserve 
the complet market hypothesis and remove the random component $dW(t)$.  
 
\subsection{The option pricing method with a modified portfolio}

We present a new portfolio in a complete but not efficient market.
The market is still assumed to be complete, in other words, there exists a
portfolio with assets to eliminate financial risk. However,
we relax the efficient market hypothesis by including the correlated
O-U process as noise for the underlying price dynamics.

Now, our portfolio is compounded by a number of calls $\Psi$ with
maturity $T$ and strike $K$, a quantity of bonds $\Phi$, and another
number of ``secondary calls" $\Psi'$, on the same asset, but with 
a different strike $K'$ and, eventually, different payoff or maturity time. 
Note that in the new portfolio there are no shares of the
underlying asset. Thus, instead of Eq.~(\ref{P-B-S}), we have 
\begin{equation}
P=\Psi C-\Psi'C'-\Phi B.
\label{portfolio}
\end{equation}
After assuming the net-zero investment, we obtain 
\begin{equation}
C=\phi B+\psi C', 
\label{callportfolio}
\end{equation}
where $\phi\equiv\Phi/\Psi$ is the number of bonds per call, and
$\psi\equiv\Psi'/\Psi$ is the number of secondary calls per call. We
proceed as before, thus the nonaticipating character of $\phi$ and $\psi$
allows us to write 
\begin{equation}
dC=\phi dB+\psi dC'
\label{dC2d}
\end{equation}
and, after using It\^o lemma~(\ref{ito}) for both $dC$ and
$dC'$, some simple manipulations yield 
\begin{eqnarray} 
\Biggl[\Bigl(C_t&+&\frac{\sigma^{2}}{2\tau}C_{VV}-
r C+(\mu+V)SC_S\Bigr) \nonumber \\
&-&\psi\Bigl(C'_t+\frac{\sigma^{2}}{2 \tau}C'_{VV}-r C'+
(\mu+V)SC'_S\Bigr)\Biggr]dt=\left(\psi C'_V-C_V\right)dV.
\label{eq}
\end{eqnarray}
This equation can be transformed to a deterministic one by equating to
zero the term multiplying the random differential $dV(t)$ given by
Eq.~(\ref{3a}). This, in 
turn, will determine the investor strategy giving the relative number of
secondary calls to be held. Thus, instead of B-S delta hedging, we will
have the ``psi hedging":
\begin{equation}
\psi=\frac{C_V}{C'_V}. 
\label{delta1}
\end{equation}
Then 
\begin{eqnarray} 
\frac{1}{C_V}\Bigl[C_t+\frac{\sigma^{2}}{2\tau}C_{VV}&-&r C
+(\mu+V)SC_S\Bigr]\nonumber\\
&=&
\frac{1}{C'_V}\Bigl[C'_t+\frac{\sigma^{2}}{2\tau}C'_{VV}
-r C'+(\mu+V)SC'_S\Bigr]. 
\label{dp2}
\end{eqnarray}
This equation proves, as otherwise expected, that the call has the same
partial differential equation independent of its maturity and strike.
This has been suggested in a more theoretical setting for any derivative
on the same asset (Bj\"{o}rk (1998)).

On the other hand, the two options $C$ and $C'$ have different strikes.
Then, analogously to the separation of variable method used in
mathematics (Mynt-U (1987)) and proceeding in a similar way to that used
in the study of SV cases, both sides of Eq.~(\ref{dp2}) are assumed to be
equal to an unknown function $\lambda(S,V,t)$ of the independent variables
$S$, $V$, and $t$. We thus have 
\begin{equation}
C_t+\frac{\sigma^2}{2\tau}C_{VV}+(\mu +V)SC_S-rC=\lambda C_V. 
\label{partial1}
\end{equation}
In the stochastic volatility literature, the arbitrary function
$\lambda(S,V,t)$ is known as the ``risk premium'' associated, in
our case, with the return velocity (Scott (1987); Heston (1993)). In the 
Appendix C we show that the risk premium $\lambda$ is given by
\begin{equation}
\lambda(S,V,t)=\frac{V}{\tau}.
\label{lambda}
\end{equation} 
A substitution of Eq.~(\ref{lambda}) 
into Eq.~(\ref{partial1}) yields a
closed partial differential for the call price $C(S,V,t)$ which is
\begin{equation}
C_t+\frac{\sigma^2}{2 \tau}C_{VV}-\frac{V}{\tau}C_V+
(\mu+V)SC_S-r C=0. 
\label{partial2}
\end{equation}

For the European call, Eq.~(\ref{partial2}) has to be solved with the
``final condition"~(\ref{finalC}) at maturity time which is 
$C(S,V,T)=\max[S(T)-K,0]$.
The solution to Eq.~(\ref{partial2}) subject to Eq.~(\ref{finalC}) is
given in Appendix D and reads 
\begin{equation}
C(S,V,t)=e^{-r (T-t)}\left[Se^{\beta(T-t,V)}N(z_1)- KN(z_2)\right], 
\label{call1}
\end{equation}
where 
$z_1=z_1(S,V,T-t)$, $z_2=z_2(S,V,T-t)$ are given by Eq.~(\ref{z12}) of 
Appendix D, and 
\[
\beta(t,V)=m(t,V)+K_{11}(t)/2,
\]
where $m(t,V)$ and $K_{11}(t)$ are given by Eqs.~(\ref{mr}) and~(\ref{k}).

The option price~(\ref{call1}) depends on both the price $S$ and the
velocity $V$ of the underlying asset at time $t$, {\it i.e.}, at the time
at which the call is bought. They are therefore the initial variables of
the problem. However, while the initial price $S$ is always known, the
initial velocity $V$ is unknown. 
The velocity is thus assumed to be in the stationary regime so that
its probability density function is as shown in Eq.~(\ref{pstat}). We
therefore average over the unknown initial velocity and define
$\overline{C}$ by 
\begin{equation}
\overline{C}(S,t)\equiv \int_{-\infty}^{\infty}C(S,V,t)p_{st}(V)dV, 
\label{barC}
\end{equation}
and from Eqs.~(\ref{pstat}) and~(\ref{call1}) we have 
\begin{equation}
\overline{C}(S,t)=e^{-r (T-t)}\left[Se^{\beta(T-t)}
N(\bar{z}_1)-KN(\bar{z}_2)\right], 
\label{call2}
\end{equation}
where
\begin{equation}
\beta(t)=\mu t+\kappa(t)/2, 
\label{barbeta}
\end{equation}
$\kappa(t)$ is the variance defined by Eq.~(\ref{varr}), and
$\bar{z}_{1,2}$ are given by Eq.~(\ref{barz}) of Appendix D. 
As mentioned above, Eq.~(\ref{call2}) cannot be our final price yet
because it still depends on the mean return rate $\mu$. This rate could
differ depending on whether $\mu$ is estimated by the seller or buyer of
the option and thereofore, in Eq.~(\ref{call2}), there are hidden
arbitrage opportunities.

Therefore, we must proceed in a similar way as in the martingale
option pricing theory of Eq.~(\ref{beta=rho}) and define the final call price,
$C_{OU}(S,t)$, as price $\overline{C}$ when $\beta(t)$ is replaced by
$rt$. That is:
\begin{equation}
C_{OU}(S,t)\equiv
\overline{C}(S,t)\Bigr|_{\beta(t)\rightarrow r t},
\label{call3}
\end{equation}
and this price completely agrees with the one derived in Section~\ref{sec:projected} (see Eq.~(\ref{call4})).

\subsection{The projected process and the modified portfolio\label{sec:mod}} 

Suppose we start from the modified portfolio~(\ref{callportfolio}) but assuming that the share price is given by the projected process~(\ref{dSe}) instead of the two-dimensional O-U process~(\ref{2a})-(\ref{3a}). In this case, one can obtain the same 
option price as before ({\it cf.} Eq.~(\ref{call4})). However, the hedging strategy will be given by the following function 
\begin{equation}
\psi(S,t)=\frac{C_S}{C'_S}.
\label{psibar}
\end{equation}
Let us prove this. We start from Eq.~(\ref{dC2d}):
\[
dC=\phi dB+\psi dC',
\]
Now, instead of Eq.~(\ref{eq}) we have (see It\^o lemma~(\ref{sde2}))
\begin{eqnarray}
\Biggl[\Bigl(C_t+\frac{1}{2}\dot{\kappa}(T-t)C_{SS}-r C\Bigr)
-\psi\Bigl(C'_t&+&\frac{1}{2}\dot{\kappa}(T-t)C'_{SS}-r C'\Bigr)\Biggr]dt
\nonumber \\
&&=\left(\psi C'_S-C_S\right)dS.
\label{sde4}
\end{eqnarray}
And the removal of risk implies Eq.~(\ref{psibar}). 
The psi hedging given by Eq.~(\ref{psibar}) is equivalent to
the psi hedging defined in Eq.~(\ref{delta1}) although 
now it is represented in terms of the final price $C_{OU}(S,t)$ instead of the
intermediate price $C(S,V,t)$. Substituting Eq.~(\ref{psibar}) into 
Eq.~(\ref{sde4}) and reasoning along the same lines as above 
(see Eq.~(\ref{partial1})) we obtain
\begin{equation}
C_t +\frac{1}{2} \dot{\kappa}(T-t)S^2C_{SS}-rC=\lambda C_S,
\label{sde5}
\end{equation}
where $\lambda=\lambda(S,t)$ is the ``risk premium'' for the effective process which is now obviously independent of the velocity $V$. Combining Eqs.~(\ref{dSe}),~(\ref{sde2}) and~(\ref{sde5}), we get
\begin{eqnarray}
dC(S,t)= \Bigl\{r C +
\Bigl[\frac{\lambda}{S}+\mu&+&\frac{1}{2}\dot{\kappa}(T-t)\Bigr]SC_S \Bigr\}dt
\nonumber \\
&&+\sqrt{\dot{\kappa}(T-t)}SC_S dW(t).
\label{sde6}
\end{eqnarray}
Hence, the conditional expected value of $dC$ reads 
\begin{equation}
E[dC|C]= \left\{rC+\left[\frac{\lambda}{S}+ \mu+\frac{1}{2}\dot{\kappa}(T-t)\right]
SC_S\right\}dt,
\label{sde7}
\end{equation}
but the equilibrium of the market implies that $E[dC|C]=rCdt$. Therefore,
\begin{equation}
\lambda=-S\left[\mu+\frac{1}{2}\dot{\kappa}(T-t)\right],
\label{sde8}
\end{equation}
and Eq.~(\ref{sde5}) reads
\[
C_t +\frac{1}{2} \dot{\kappa}(T-t)S^2C_{SS}-rC +
\left[\mu+\frac{1}{2}\dot{\kappa}(T-t)\right] SC_S=0,
\]
Finally, the absence of arbitrage opportunities requires the replacement (see 
Eq.~(\ref{barbeta})) 
\[
\mu+\dot{\kappa}(T-t)/2 \longrightarrow r.
\]
Thus, the option price equation is
\begin{equation}
C_t=rC -rSC_S-\frac{1}{2} \dot{\kappa}(T-t)S^2C_{SS},
\label{sde9}
\end{equation}
which agrees with Eq.~(\ref{sde3}). 

Note that both procedures, the original B-S method presented in Section~\ref{sec:projected}
and our method, result in the same partial differential equation for the call 
price. However, each method uses a different hedging strategy because they 
start from a different portfolio. 

\section{The call price by the equivalent martingale measure method\label{sec:martingale}}

As was shown by Harrison and Kreps (1979) and Harrison and Pliska (1981),
the B-S option price can also be found using martingale methods. This is
a shorter, although more abstract way, to derive an expression for the
call price. The main advantage is that one only needs to know the
probability density function governing market evolution which, in turn,
allows one to obtain an option price in situations where B-S assumptions
are not applicable. The drawback is that one is not sure of whether the
price obtained by martingale methods is the fair price of the call because
of the omission of arbitrage and hedging.

We will now show that, in the present case, the price obtained by
martingale methods completely agrees with our extended B-S
price~(\ref{call4}). The equivalent martingale measure theory imposes 
the condition that, in a ``risk-neutral world", the stock price $S(t)$ 
evolves, on average, as a riskless bond (Harrison and Pliska (1981)). 

Let $p^*(S,t|S_0,t_0)$ be the equivalent martingale measure associated
with asset price $S(t)$ conditioned on $S(t_0)=S_0$. Define the martingale 
conditional expected value 
\[
E^*\left[S(t)|S_0\right]=\int_{0}^{\infty}Sp^*(S,t|S_0,t_0)dS. 
\]
Then the risk-neutral assumption requires that
\[
E^*\left[S(t)|S_0\right]=S_0
e^{r (t-t_0)},
\]
where $r$ is the constant spot interest rate.
On the other hand, 
\[
E\left[S(t)|S_0\right]=\int_{0}^{\infty}Sp(S,t|S_0,t_0)dS.
\]
Assuming that the initial velocity is in the statioanry regime, the 
marginal density $p(S,t|S_,t_0)$ is given by Eq.~(\ref{pSa}) of 
Appendix A. Therefore,
\[
E\left[S(t)|S_0\right]=S_0\exp\left[\beta(t-t_0)\right],
\]
where $\beta(t)=\mu t+\kappa(t)/2$ with $\kappa(t)$ given by 
Eq.~(\ref{varr}). We thus see that the equivalent
martingale measure is accomplished by the replacement
\begin{equation}
\beta(t)\longrightarrow r t.
\label{beta=rho}
\end{equation}
In consequence,
\begin{eqnarray}
p^*(S,t|S_0,t_0)=&&\frac{1}{S\sqrt{2\pi\kappa(t-t_0)}}
\nonumber \\
&& \times \ \exp\left\{-\frac{[\ln(S/S_0)-r (t-t_0)+\kappa(t-t_0)/2]^2}
{2\kappa(t-t_0)}\right\},
\label{martingale2} 
\end{eqnarray}
which is the so called ``the risk-neutral pdf" for the stock price and it
is a consequence of the absence of arbitrage demmand. Now, it is possible
to express the price for the European call option by defining its value as
the discounted expected gain due to holding the call. That is (Harrison
and Pliska (1981)), 
\begin{eqnarray}
C^*(S,t)&=&e^{-r (T-t)}
E^*\bigl\{\max[S(T)-K,0]|S(t)=S\bigr\}\nonumber\\ 
&=&e^{-r (T-t)}\int_K^\infty(S'-K)p^*(S',T|S,t)dS',
\label{martingale3}
\end{eqnarray}
and the final result for the call is obtained by calculating the expected
value with the equivalent martingale measure defined in
Eq.~(\ref{martingale2}). 
The martingale price agrees exactly with our previous price in
Eq.~(\ref{call4}), $C^*(S,t)=C_{OU}(S,t)$. We can thus say that, in the
O-U case, both option pricing methods are completely equivalent although
martingale theory does not require the construction of a portfolio and
ignores any hedging strategy.

\section{Greeks and Hedging\label{sec:hedging}}

We briefly derive the Greeks for the O-U case. Since the O-U call price
has the same functional form as the B-S price but replaces $\sigma^2(T-t)$ 
by $\kappa(T-t)$, the O-U Greeks will have the same functional
form as B-S Greeks with the same replacement except for Vega, 
${\cal V}=\partial C/\partial\sigma$, and $\theta=\partial C/\partial t$.
Thus, for $\delta=\partial C/\partial S$, $\gamma=\partial^2C/\partial
S^2$, and $\rho=\partial C/\partial r$, we have (Hull (2000))
\begin{eqnarray}
{\delta}_{OU}=N(d_1^{OU}), \quad
&&{\gamma}_{OU}=\frac{e^{-(d_1^{OU})^2/2}}{S\sqrt{2\pi\kappa(T-t)}}, \nonumber
\\
&&{\rho}_{OU}=K(T-t)e^{-r (T-t)}N(d_2^{OU}).
\label{Greeks}
\end{eqnarray}
Since $d_{1,2}^{OU}\geq d_{1,2}^{BS}$ for all $S$ and $t$ and $N(z)$ is a
monotone increasing function, we see that ${\delta}_{OU}\geq{\delta}_{BS}$
and ${\rho}_{OU}\geq{\rho}_{BS}$. Hence, the O-U call price is more
sensitive to changes in stock price and interest rate than the B-S
price. 

On the other hand, from Eq.~(\ref{call4}) and taking into account the
identity 
\begin{equation}
SN'(d_1)-Ke^{-r (T-t)}N'(d_2)=0,
\label{identity}
\end{equation}
we have 
\begin{equation}
{\cal V}_{OU}=(S/\sigma)[\kappa(T-t)/2\pi]^{1/2}e^{-(d^{OU}_1)^2/2}, 
\label{vega}
\end{equation}
and
\begin{equation}
{\theta}_{OU}=-Ke^{-r (T-t)}\left[r N(d_2^{OU})+
\frac{\sigma^2 (1-e^{-(T-t)/\tau})}{2\sqrt{2\pi\kappa(T-t)}}
\ e^{-(d^{OU}_2)^2/2}\right].
\label{theta}
\end{equation}
Since $d_1^{OU}\geq d_1^{BS}$, one can easily see that ${\cal V}_{OU}\leq{\cal V}_{BS}$ for all values of $S/K$, $T-t$ and $\tau$. Thus
our correlated call price is less sensitive to any change of underlying
volatility $\sigma$ than is the B-S price. 

We conclude with the psi hedging. For the two-dimensional O-U case the hedging
strategy is given by the function $\psi(S,V,t)$ specifying the number
of secondary calls to be hold. However, the hedging given by Eq.~(\ref{delta1})
depends on the velocity $V$ and is not expressed in terms of the final call 
price $C_{OU}=C(S,t)$. As we have shown in Section~\ref{sec:mod}, 
psi hedging in terms of $C_{OU}$ can only be derived from the effective 
one-dimensional process~(\ref{dSe}). In this case, the removal of the 
randomness coming from $dS$ implies that hedging is given by 
Eq.~(\ref{psibar}). Since $C_S=\delta_{OU}$, we see from Eqs.~(\ref{psibar}) 
and Eq.~(\ref{Greeks}) that
\begin{equation}
\psi(S,t)=\frac{N(d_1)}{N(d'_1)}.
\label{psi2b}
\end{equation}

\begin{figure}[tbp]
\begin{center}
\includegraphics[angle=-90,scale=0.55]{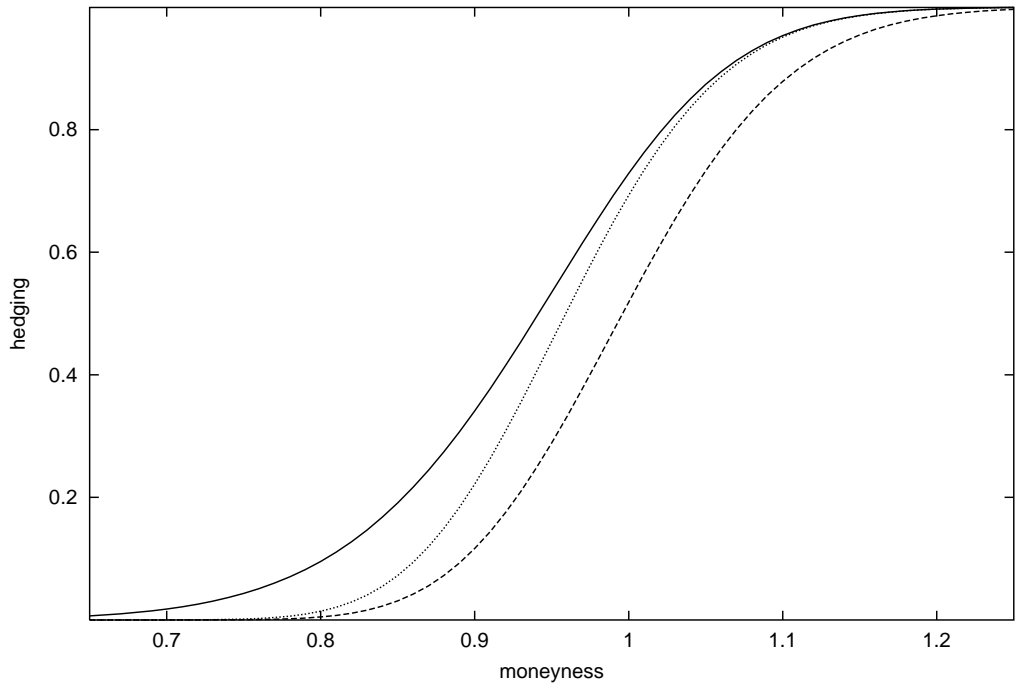}
\end{center}
\caption{Hedging in terms of the moneyness. Psi hedging and delta hedging as a function of the moneyness $S/K$. The solid line represents psi hedging when $\tau=1$ day, the time to expiration is $T-t=20$ days, and the exercising price of the secondary call is $K'=0.9 K$. The dotted line corresponds to the delta hedging still assuming the O-U asset model with the same correlation and expiration time. The dashed line corresponds to B-S delta hedging ($r$ and $\sigma$ as in Fig. 2).}
\label{fig7}
\end{figure}

Now, we take the secondary option to be an European call with maturity $T$ and exercising price $K'<K$, where $T$ and $K$ refer to the primary option. We plot in Fig.~\ref{fig7} the psi hedging as a function of the moneyness. We see there that the $\psi$ hedging is always greater than $\delta_{OU}$ and $\delta_{BS}$ hedgings. Since $N(d'_1) \rightarrow 1$ when $K'\rightarrow 0$, the psi hedging approaches to the delta hedging $\delta_{OU}$ as the moneyness of the secondary call tends to infinity. This is consistent with the fact that secondary calls have the same price as the underlying stock when its exercising price is zero , {\it i.e.} $C'\rightarrow S$ as  $K'\rightarrow 0$
(see Eq.~(\ref{call4})).  
Therefore, having secondary calls with exercising price equal to zero 
is equivalent to own underlying shares and the O-U psi hedging coincides with 
the O-U delta hedging.

\begin{figure}
\begin{center}
\includegraphics[angle=-90,scale=0.55]{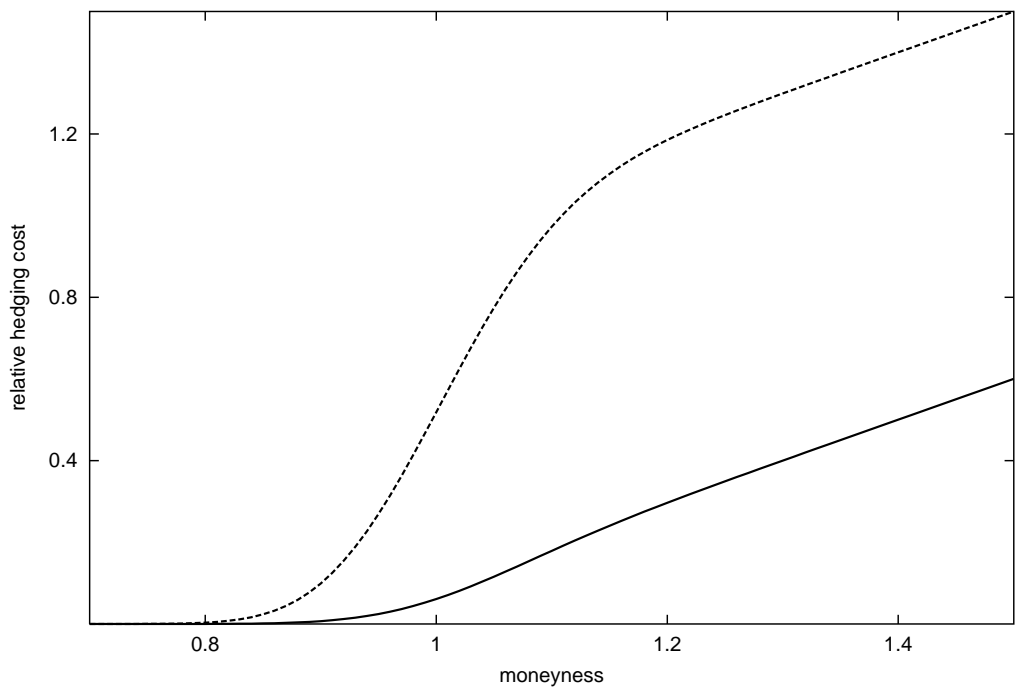}
\end{center}
\caption{Relative hedging costs $\psi C'/K$ and $\delta S/K$ as a function of the moneyness $S/K$. The solid line represent psi hedging cost when $\tau=1$ day, the time to expiration is $T-t=20$ days, and the exercising price of the secondary call is $K'=0.9 K$. The dotted line corresponds to the delta hedging with $\tau=1$ day and $T-t=20$ days ($r$ and $\sigma$ as in Fig. 2).}
\label{fig8}
\end{figure}

As we have mentioned, psi hedging $\psi$ indicates the number of secondary 
calls per call to be hold if we follow a risk-free strategy with the 
modified portfolio~(\ref{portfolio}). Therefore, the money invested
to carry out this strategy is given by $\psi C'$. That is 
\begin{equation}
\psi C'= \frac{N(d_1)}{N(d'_1)} \left[S N(d'_1)-K' e^{-r(T-t)} N(d'_2)\right],
\label{psic}
\end{equation}
where we have combined the Eqs.~(\ref{call4}) and~(\ref{psi2b}). On the other
hand, delta hedging also indicates the number of shares per call to be hold
in a risk-free strategy with the B-S portfolio~(\ref{P-B-S}). And, 
analogously, the money necessary to perform this strategy is
\begin{equation}
\delta S= S N(d_1),
\label{deltas}
\end{equation}
where $\delta$ is given by Eq.~(\ref{Greeks}). We compare these quantities 
in order to know which hedging is cheaper for the investor. From 
Eqs.~(\ref{psic}) and~(\ref{deltas}), we see
\[
\frac{\psi C'}{\delta S}= 1-\frac{ K'e^{-r(T-t)} \ N(d'_2)}{S\ N(d'_1)},
\]
but\footnote{This is straightforward to prove from Eq.~(\ref{psic}) since $\psi C'\geq0$.}
\[
0 \leq \frac{ K'e^{-r(T-t)} \ N(d'_2)}{S\ N(d'_1)} \leq 1.
\]
Therefore, $\psi C'<\delta S$ and psi hedging is always less
expensive than delta hedging. Note that when $K'\rightarrow 0$ both 
strategies have the same cost. In Fig.~\ref{fig8} we plot, as a function of
moneyness, the relative psi hedging cost, $\psi C'/K$, along with the relative 
delta hedging cost, $\delta S/K$. We see there that psi hedging 
is considerably less expensive than delta hedging and this difference 
increases with moneyness. Indeed, for an ATM call ($S/K=1.00$) 
and with parameter values as that of Fig.~\ref{fig8}, delta 
hedging is approximately $800\%$ more expensive than psi hedging.

Combining Eqs.~(\ref{call4B-S}),~(\ref{psibar}) 
and~(\ref{identity}) one can easily show that when $\tau=0$ the O-U psi 
hedging is $\psi_{BS}=N(d_1^{BS})/N(d_1^{BS'})$
\footnote{We use the subscript $BS$ in $\psi_{BS}$ to indicate that this 
hedging refers to an uncorrelated stock ($\tau=0$), as in the B-S world}, 
where the prime refers to the secondary call. Since $\delta=C_S=N(d_1)$, we 
have
\[
\psi_{BS}=\frac{\delta_{BS}}{\delta'_{BS}}.
\]
Finally, for the secondary call, whose exercising price goes to zero, 
$\delta'_{BS}\rightarrow 1$ and, again, B-S psi hedging and B-S delta hedging 
coincide. 

\section{Conclusions\label{sec:conc}}

We have developed option pricing with perfect hedging in an inefficient
market model. The inefficiency of the market is related to the fact that
the underlying price variations are autocorrelated over an arbitrary time
period $\tau$. In order to take these correlations into account we have 
modelled the underlying price $S(t)$ as a singular diffusion process in two
dimensions (O-U process) instead of the standard assumption that $S(t)$ is
a one-dimensional diffusion given by the geometric Brownian motion with 
constant volatility. 

The option pricing method has been developed by keeping perfect hedging
with a riskless strategy which finally results in a closed and exact
expression for the European call. Our pricing formula has the same
functional form as the B-S price but replaces the variance of the Wiener
process by the variance of the O-U process. The O-U variance, $\kappa(t)$,
is smaller than the B-S variance, $\sigma^2t$, which implies that the
equivalent volatility in the O-U case is lower than B-S
volatility\footnote{Since the volatility $\sigma$ is the square root of
the variance per unit time, one can define, in the O-U case, an equivalent
volatility by $\sigma_{OU}=\sqrt{\dot{\kappa}(t)}$, where the dot denotes
time derivative. From Eq.~(\ref{varr}) we see that
$\sigma_{OU}/\sigma=\sqrt{1-e^{-t/\tau}}\leq 1$.}. But less volatility
implies a lower option price. We have indeed proved that the B-S call
price is always greater than the O-U price. In other words, the assumption
of uncorrelated assets overprices the European call. This agrees with the
fact that correlation, which can be regarded as a form of predictability,
implies less risk and therefore a lower price for the option. We have
quantified this overprice and showed that B-S formula notably overprices 
options and, more strikingly, that the overprice persists for a long time
regardless of the strength of correlations. We have also analyzed the 
sensitivity of the O-U price to several conditions. Thus we have proved 
that while $C_{OU}$ is more sensitive to changes in the interest rate and 
stock price than $C_{BS}$, it is also less sensitive to any change of the 
volatility. The practical consequences of
this are nontrivial. 

The option price and the hedging strategy have been obtained using two 
different approaches. The most straightforward way of getting the call price
is by means of a projection onto a one-dimensional process with a 
time-varying volatility.
A second way of obtaining the option price starts with the complete
two-dimensional O-U process~(\ref{2a})-(\ref{3a}). This is a longer procedure 
but opens the door to a new hedging strategy: the psi hedging. We have 
therefore two ways of acheiving the perfect hedging: the usual one consisting in 
holding underlying assets (delta hedging), and the second one which uses 
secondary calls\footnote{The construction of the portfolio
with secondary calls is one simple way of proceeding. Obviously,
any other secondary derivative on the same asset would serve.}
instead of assets (psi hedging). We have shown that this last strategy can be considerable less expensive than the delta hedging and can avoid a possible lack of liquidity of underlying shares. Finally, the proportion of secondary calls to be held, {\it i.e.}, the psi hedging, converges towards O-U delta hedging when the exercising price of the secondary call tends to zero.

In practice our method of valuation requires the estimate of one more
parameter, the correlation time, than in the B-S Wiener case. 
Assuming that the underlying
asset is driven by O-U noise one can find an estimate for the correlation
time $\tau$ by evaluating the variance $\kappa(t)$ of the asset return.
Once one has an estimate of this variance the correlation time is given in
Eq.~(\ref{varr}). 

We finally mention that one interesting extension of the valuation method
presented is to the American option. Although this case is more involved,
one is probably able to obtain, at least an approximate or a numerical
result using a combination of first passage times and martingale methods,
as recently presented by Bunch and Johnson (2000). In any case we
believe that the effects of autocorrelations on the valuation of an
American
option will be even more critical than for the European call. This case is
under present investigation. 

\ack

The authors acknowledge helpful comments and discussions with Alan McKane, Miquel Montero, Josep M. Porr\`a and Jaume Puig. We are particularly grateful to Santiago Carrillo and George H. Weiss for their many suggestions to improve the manuscript. 
This work has been supported in part by Direcci\'on General de Proyectos de Investigaci\'on under contract No. BFM2000-0795, and by Generalitat de Catalunya 
under contract No. 2000 SGR-00023.

\appendix 

\section{Mathematical properties of the model}

We present some of the most important properties of the model given by the
pair of stochastic equations in Eqs.~(\ref{4a}) and~(\ref{4b}). Their formal 
solutions are
\[
V(t)= V_0 e^{-(t-t_0)/\tau} +
\frac{\sigma}{\tau}\int_{t_0}^{t}e^{-(t-t')/\tau}dW(t^{\prime}), 
\]
and
\begin{eqnarray}
R(t)=\mu (t-t_0)+ V_0 \tau(1&-&e^{-(t-t_0)/\tau})
\nonumber \\&+&\frac{\sigma}{\tau}\int_{t_0}^{t}dt'
\int_{t_0}^{t'}e^{-(t'-t^{\prime\prime})/\tau}
dW(t^{\prime\prime}), 
\label{r2}
\end{eqnarray}
where we have assumed that the process begun at time $t_0$ with initial velocity
$V_0$ and return $R_0=0$.
The return $R(t)$ has the following conditional mean value
\[
E[R(t)|V_0]=\mu (t-t_0)+\tau(1-e^{-(t-t_0)/\tau})V_0,
\]
and variance
\[
\rm{Var}[R(t)|V_0]=\sigma^2\Bigl[(t-t_0)-2\tau\left(1-e^{-(t-t_0)/\tau}\right)
+ \frac{\tau}{2}\left(1-e^{-2(t-t_0)/\tau}\right)\Bigr].
\]

Since $(R(t),V(t))$ is a diffusion process in two dimensions, its joint
density $p(R,V,t)$ satisfies the following Fokker-Planck
equation (Gardiner (1985))  
\begin{equation}
p_t=-(\mu+V)p_R+\frac{V}{\tau}p_V+\frac{\sigma^2}{2\tau^2}p_{VV}. 
\label{fp1}
\end{equation}
This is to be solved subject to the initial conditions $R(t_0)=0$
and $V(t_0)=V_0$, that is
\begin{equation}
p(R,V,t_0|V_0,t_0)=\delta(R)\delta(V-V_0).
\label{initial0}
\end{equation}

A first step towards solving the problem~(\ref{fp1})-(\ref{initial0}) is the
definition of the joint Fourier transform
\[
\tilde{p}(\alpha,\beta,t)=
\int_{-\infty}^{\infty}dRe^{i\alpha R}\int_{-\infty}^{\infty}dV
e^{i\beta V}p(R,V,t).
\]
Then problem~(\ref{fp1})-(\ref{initial0}) becomes
\begin{equation}
\partial_t\tilde{p}=i\alpha\mu\tilde{p}+(\alpha-\beta/\tau)\alpha
\partial_{\beta}\tilde{p}-(\sigma^2/2\tau^2)\beta^2\tilde{p},
\label{3}
\end{equation}
\begin{equation}
\tilde{p}(\alpha,\beta,t=0)=e^{i\beta V_0}.
\label{4}
\end{equation}
We look for a solution of the form
\begin{eqnarray}
\tilde{p}(\alpha,\beta,t)=&& \exp\{i[\alpha m_1(t)+\beta m_2(t)]\} \nonumber \\
&&\times \exp\{-[K_{11}(t)\alpha^2+K_{12}(t)\alpha\beta+K_{22}(t)\beta^2]/2\},
\label{5}
\end{eqnarray}
where $m_i(t)$ and $K_{ij}(t)$ are functions to be determined. 
We substitute Eq.~(\ref{5}) into~(\ref{3}) and identify term by term. 
We have
\begin{eqnarray*}
\dot{m}_1 = \mu + m_2, &&\quad \dot{m}_2= -m_2/\tau;
\\
\dot{K}_{22}+(2/\tau)K_{22}=\sigma^2\tau^2,\quad
&& \dot{K}_{12}+(1/\tau)K_{12}=2K_{22}(t),\quad\dot{K}_{11}=2K_{12},
\end{eqnarray*}
with the intial conditions, according to Eqs.~(\ref{4})-(\ref{5}), given by
\[
m_2(0)=V_0, \quad m_1(0)=K_{ij}(0)=0\quad(i,j=1,2).
\]
The solution reads
\[
m_1(t)=\mu t+V_0\tau\left(1-e^{-t/\tau}\right),\quad m_2(t)=V_0e^{-t/\tau},
\]
and $K_{ij}(t)$ are given by
\begin{equation}
K_{11}(t)=\sigma^2\left[t-2\tau\left(1-e^{-t/\tau}\right)+
\frac{\tau}{2}\left(1-e^{-2t/\tau}\right)\right], 
\label{k11}
\end{equation}
\begin{equation}
K_{12}(t)=\frac{\sigma^2}{2}\left(1-e^{-t/\tau}\right)^2, 
\qquad K_{22}(t)=\frac{\sigma^2}{2\tau}\left(1-e^{-2t/\tau}\right),
\label{k22}
\end{equation}

The inverse Fourier transform of Eq.~(\ref{5}) yields the Gaussian density 
\begin{eqnarray}
p&(&R,V,t|V_0,t_0)=\frac{1}{2\pi\sqrt{\det[\mbox{\boldmath$K$}(t-t_0)]}}
\exp\Biggl\{-\frac{\bigl(V-V_0e^{-(t-t_0)/\tau}\bigr)^2}{2K_{22}(t-t_0)}
\nonumber\\
&&-\frac{\left[K_{22}(t-t_0)(R-m(t-t_0,V_0))-
K_{11}(t-t_0)\left(V-V_0e^{-(t-t_0)/\tau}\right)\right]^2}
{2K_{22}(t-t_0)\det[\mbox{\boldmath$K$}(t-t_0)]}\Biggr\},
\nonumber \\
\label{p1}
\end{eqnarray}
where 
\begin{equation}
\det[\mbox{\boldmath$K$}(t)]\equiv K_{11}(t)K_{22}(t)-K_{12}^2(t).
\label{bfDelta}
\end{equation}
and 
\begin{equation}
m(t,V_0)=\mu t +V_0\tau\left(1-e^{-t/\tau}\right). 
\label{m0}
\end{equation}
Notice that the joint density~(\ref{p1}) is a function of the time differences
$t-t_0$ 
where $t_0$ is the initial observation time, so that the two-dimensional 
diffusion $(S(t),V(t))$ is a time homogeneous process and, without loss of 
generality, we may assume that $t_0=0$. 

The marginal pdf of the velocity $V(t)$, 
\[
p(V,t|V_0)=\int_{-\infty}^{\infty}p(R,V,t|V_0)dR, 
\]
is 
\begin{equation}
p(V,t|V_0)=\frac{1}{\sqrt{2\pi K_{22}(t)}} \exp\left[-\frac{
\left(V-V_0e^{-t/\tau}\right)^2}{2K_{22}(t)}\right]. 
\label{pv0}
\end{equation}
In the stationary regime ($t\rightarrow\infty$) we find a normal density
independent of the initial velocity: 
\begin{equation}
p_{st}(V)=\frac{1}{\sqrt{\pi(\sigma^2/\tau)}}
e^{-\tau V^2/\sigma^2}. 
\label{pstatapp}
\end{equation}
Analogously, the marginal density of the return $R(t)$,
\[
p(R,t|V_0)=\int_{-\infty}^{\infty}p(R,V,t|V_0)dV, 
\]
is 
\begin{equation}
p(R,t|V_0)=\frac{1}{\sqrt{2\pi K_{11}(t)}}
\exp\left\{-\frac{\left[R-m(t,V_0)\right]^2}{2K_{11}(t)}\right\}. 
\label{pr0}
\end{equation}

If we assume that the initial velocity $V_0=V(0)$ is a random
variable distributed according to the pdf in Eq.~(\ref{pstatapp}). We can
therefore average the above densities to obtain a pdf independent of $V_0$.
That is, 
\[
p(R,V,t)=\int_{-\infty}^{\infty}p(R,V,t|V_0)p_{st}(V_0)dV_0, 
\]
and similarly for the marginal pdf's $p(R,t)$ and $p(V,t)$. Since we are
mainly interested on the marginal distribution of the return we will give its
explicit expression. Thus, from Eqs.~(\ref{pstatapp}) and~(\ref{pr0}) we have 
\begin{equation}
p(R,t)=\frac{1}{\sqrt{2\pi\kappa(t)}}
\exp\left[-\frac{\left(R-\mu t\right)^2}{2\kappa(t)}\right], 
\label{prapp}
\end{equation}
where $\kappa(t)$ is given by Eq.~(\ref{varr}).
Alternatively, the distribution of the underlying price $S=S_0e^{R}$
is given by the log-normal density 
\begin{equation}
p(S,t|S_0)=\frac{1}{S\sqrt{2\pi\kappa(t)}}
\exp\left[-\frac{\left(\ln S/S_0-\mu t\right)^2}{2\kappa(t)}\right]. 
\label{pSa}
\end{equation}
From this we easily see that the conditional probability $p(S',T|S,t)$ when
$t\leq T$ is
\begin{equation}
p(S',T|S,t)=\frac{1}{S'\sqrt{2\pi\kappa(T-t)}}
\exp\left[-\frac{\left[\ln S'/S-\mu (T-t)\right]^2}{2\kappa(T-t)}\right].
\label{pS}
\end{equation}

\section{The It\^o formula for processes driven by O-U noise}

In this Appendix we generalize the It\^o formula for
processes driven by Ornstein-Uhlenbeck noise. This is applied to the share
price $S(t)$ which is governed by the pair of stochastic 
equations~(\ref{2a})-(\ref{3a})
\begin{equation}
dS(t)=S(\mu+V)dt,
\qquad
dV(t)=-\frac{V}{\tau}dt+\frac{\sigma}{\tau}dW.
\label{a2}
\end{equation}

Consider a generic function $f(S,V,t)$ which depends on all of the
variables that characterize the underlying asset. The differential of
$f(S,V,t)$ is defined by 
\begin{equation}
df(S,V,t)\equiv f\bigl(S(t+dt),V(t+dt),t+dt\bigr)- 
f\bigl(S(t),V(t),t\bigr). 
\label{a1}
\end{equation}
But the Taylor expansion of~(\ref{a1}) yields 
\begin{eqnarray}
df(S,V,t)=f_S dS&+&f_V dV+f_t dt\nonumber \\
&&+\frac{1}{2}f_{SS}dS^2+
\frac{1}{2}f_{VV}dV^2 +f_{SV}dSdV+\cdots, 
\label{df}
\end{eqnarray}
where the expansion also involves higher order differentials such as 
$(dt)^2$, $(dS)^3$, $(dV)^3$, etc.
However, the differential of the Wiener process, $dW$, satisfies
the well-known property, in the mean-square sense, 
$dW(t)^2= dt$ (Gardiner (1985)). And from the pair of equations~(\ref{a2}) we then
see that $dS^2$ is of order $dt^2$ while $dV^2$ is of order $dt$ and
$dSdV$ is of order $dt^{3/2}$. Therefore, up to order $dt$, Eq.~(\ref{df})
reads 
\begin{equation}
df(S,V,t)= f_S dS+f_V dV+f_t dt+\frac{\sigma^2}{2\tau^2}f_{VV}dt,
\label{a4}
\end{equation}
which is the It\^o formula for our singular two-dimensional
process~(\ref{2a})-(\ref{3a}).

Suppose now we start from the effective one-dimensional SDE~(\ref{dRe})
\begin{equation}
dR(t)=\mu dt +\sqrt{\dot{\kappa}(T-t)}dW(t).
\label{a5}
\end{equation}
We will prove that the corresponding SDE for the stock price defined as $S=S_0 e^R$ is
given by Eq.~(\ref{dSe}). In effect, substituting Eq.~(\ref{a5}) in the Taylor expansion
\[
dS(R)=S_R dR+\frac{1}{2}S_{RR}dR^2+\cdots,
\]
neglecting orders higher than $dt$ and taking into account that 
$dR^2=\dot{\kappa}(T-t)dt$ (in mean square sense), we finally obtain
\begin{equation}
\frac{dS(t)}{S(t)}=\left[\mu+\dot{\kappa}(T-t)/2\right]dt+
\sqrt{\dot{\kappa}(T-t)}dW(t),
\label{itose1}
\end{equation}
which is Eq.~(\ref{dSe}).

Moreover, we can also give the differential of a generic function
$f(S,t)$ when underlying obeys SDE~(\ref{itose1}). In this case, we have 
\begin{equation}
df(S,t)= f_S dS+f_t dt+\frac{1}{2}\dot{\kappa}(T-t) S^2f_{SS}dt,
\label{itose}
\end{equation}
where again we have neglected higher order contributions than $dt$.

\section{A derivation of the risk premium}

We proceed to find a closed expression for the arbitrary function
$\lambda(S,V,t)$ that appears in Eq.~(\ref{partial1}). The call price $C$ 
is a function of $S,V$, and $t$. We now consider this
function taking into account that $S=S(t)$ and $V=V(t)$ follow
Eqs.~(\ref{2a}) and~(\ref{3a}), respectively.
This therefore allows us to evaluate the random differential $dC$ using
the It\^o lemma, as a result we find that
\[
dC=\left[C_t+(\mu+V)SC_S+\frac{\sigma^2}{2\tau}C_{VV}\right]dt+C_VdV.
\]
After using Eqs.~(\ref{partial1}) and~(\ref{3a}), we have
\begin{equation}
dC=\left[r C+\left(\lambda-\frac{V}{\tau}\right)C_V\right]dt+ 
\frac{\sigma}{\tau}C_VdW. 
\label{40}
\end{equation}
The expected value of $dC$, on the assumption that $C(t)=C$ is known,
reads 
\begin{equation}
E[dC|C]=\left[r C+
\left(\lambda-\frac{V}{\tau}\right)C_V\right]dt. 
\label{lambda1}
\end{equation}
We claim that this average must grow at the same rate as the risk-free
bond:
\begin{equation}
E[dC|C]=r C dt, 
\label{lambda2}
\end{equation}
since otherwise the option would not be in equilibrium (Hull (2000)). 
In some sense, this assumption is similar to that of the 
equivalent martingale measure demand expecting that markets grow in average 
as the risk-free bond (Harrison and Pliska (1981)).
 
The substitution of Eq.~(\ref{lambda2})
into Eq.~(\ref{lambda1}) yields the following expression for the risk
premium $\lambda(S,V,t)$: 
\begin{equation}
\lambda=\frac{V}{\tau}. 
\label{lambda3}
\end{equation}

\section{Solution to the problem in Eqs.~(\ref{partial2})-(\ref{finalC})}

We will solve Eq.~(\ref{partial2}) subject to the final
condition in Eq.~(\ref{finalC}). Define a new independent variable $Z$  
\begin{equation}
S=e^Z,
\label{b0}
\end{equation}
where the domain of $Z$ is unrestricted. The problem posed in
Eqs.~(\ref{partial2})-(\ref{finalC}) now reads 
\begin{equation}
C_t=rC-(\mu+V)C_Z+\frac{V}{\tau}C_V-\frac{\sigma^2}{2 \tau}C_{VV}, 
\label{b1}
\end{equation}
\begin{equation}
C(Z,V,T)=\max[{e^Z-K,0}]. 
\label{b2}
\end{equation}
The solution to this problem can be written in the form 
\begin{equation}
C(Z,V,t)=\int_{-\infty}^{\infty}dZ'
\int_{-\infty}^{\infty}dV'\max[e^{Z'}-K]
G(Z,V,t|Z',V',T), 
\label{b3}
\end{equation}
where $G(Z,V,t|Z',V',T)$ is the Green function for the
problem (Mynt-U (1987)), {\it i.e.}, $G(Z,V,t|Z',V',T)$ is the solution to 
\begin{equation}
G_t=rG-(\mu+V)G_Z+\frac{V}{\tau}G_V-\frac{\sigma^2}{2 \tau}G_{VV}, 
\label{b4}
\end{equation}
with the final condition
\begin{equation}
G(Z,V,T|Z',V',T)=\delta(Z-Z')\delta(V-V'), 
\label{b5}
\end{equation}
where $\delta(X-X')$ is the Dirac delta function. Define
$\overline{G}=e^{-rt}G$, then the final-value problem in Eqs.~(\ref{b4})
and~(\ref{b5}) reads
\begin{equation}
\overline{G}_t=-(\mu+V)\overline{G}_Z+\frac{V}{\tau}\overline{G}_V-
\frac{\sigma^2}{2 \tau}\overline{G}_{VV}, 
\label{bb4}
\end{equation}
\begin{equation}
\overline{G}(Z,V,T|Z',V',T)=e^{-rT}\delta(Z-Z')\delta(V-V').
\label{bb5}
\end{equation}
Note that Eq.~(\ref{bb4}) is the backward equation corresponding to
Eq.~(\ref{fp1}). Therefore, Eq.(\ref{p1}) permits us to write the
solution to the problem posed in
Eqs.~(\ref{bb4})-(\ref{bb5}) (Gardiner (1985)). This solution implies that
$G$ is
\begin{eqnarray}
\fl G(Z,V,t|Z',V',T)=\frac{1}{2\pi\sqrt{\det[\mbox{\boldmath$K$}(T-t)}]}
\exp\Biggl\{-r(T-t)-
\frac{\left[V'-Ve^{-(T-t)/\tau}\right]^2}{2K_{22}(T-t)}
\nonumber\\
\lo -\frac{\left[K_{22}(T-t)\left(Z'-Z+m(T-t,V)\right)-
K_{11}(T-t)\left(V'-Ve^{-(T-t)/\tau}\right)\right]^2}
{2K_{22}(T-t)\det[\mbox{\boldmath$K$}(T-t)]}\Biggr\},
\nonumber \\
\label{b9}
\end{eqnarray}
where $\det[\mbox{\boldmath$K$}(t)]$, $K_{ij}(t)$, and $m(t,V)$ are
defined in Eqs.~(\ref{bfDelta})-(\ref{m0}).

Substituting Eq.~(\ref{b9}) into Eq.~(\ref{b3}) and finally reverting to
the original variables we obtain Eq.~(\ref{call1}) with 
\begin{eqnarray}
z_1=\frac{\ln(S/K)+m(T-t,V)+K_{11}(T-t)}{\sqrt{K_{11}(T-t)}},\quad
z_2=z_1-\sqrt{K_{11}(T-t)}.
\nonumber \\ 
\label{z12}
\end{eqnarray}
Finally it can be shown, after some lengthy but simple manipulations, that
the functions $\bar{z}_{1,2}=\bar{z}_{1,2}(S,T-t)$ appearing in the
averaged price $\overline{C}(S,t)$, Eq.~(\ref{call2}), are given by
\begin{equation}
\bar{z}_1=
\frac{\ln(S/K)+\mu(T-t)+\kappa(T-t)}{\sqrt{\kappa(T-t)}},\qquad 
\bar{z}_2=\bar{z}_1-\sqrt{\kappa(T-t)},
\label{barz}
\end{equation}
where $\kappa(t)$ is given in Eq.~(\ref{varr}).

\References
%\section*{References}

%\noindent
%\hspace{0.70 cm}
%\parbox{14.5cm}{

\item [] Aurell, E., R. Baviera, O. Hammarlid, M. Serva, and A.
Vulpiani, 2000, A General Methodology to Price and Hedge Derivatives in
Incomplete Markets, {\it International Journal of  Theoretical and Applied
Finance} {\bf 3}, 1-24.

\item [] Black, F., and M. Scholes, 1973,  The Pricing of Options and Corporate Liabilities, {\it Journal of Political Economy} {\bf 81}, 637-659.

\item [] Bj\"{o}rk, T., 1998, {\it Arbitrage Theory in Continuous Time} (Oxford University Press, Oxford, UK).

\item [] Bouchaud, J. P., and M. Potters, 2000, {\it Theory of Financial Risks} (Cambridge University Press, Cambridge, UK).

\item [] Breen, W., and R. Jagannathan, 1989, Economics significance of predictable variations in stock index returns, {\it Journal of Finance} {\bf 44}, 1177-1189.

\item [] Bunch, D. S., and H. Johnson, 2000, The American Put Option and Its Critical Stock Price, {\it Journal of Finance} {\bf 55}, 2333-2356.

\item [] Campbell, J., and Y. Hamao, 1992, Predictable stock
returns in the United States and Japan: A study of long-term capital
market integration, {\it Journal of  Finance} {\bf 47}, 43-70.

\item [] Cootner, P. H., 1964, {\it The Random Character of Stock Market Prices}, (M.I.T. Press, Cambridge, MA).

\item [] Cox, J. C., and S. A. Ross, 1976, The Valuation of Options for Alternative Stochastic Processes, {\it Journal of  Financial Economics} {\bf 3}, 145-166.

\item [] Doob, J. L., 1942, The Brownian Movement and Stochastic Equations, {\it Annals of Mathematics} 43, 351-369.

\item [] Dumas, B., J. Fleming, and R. Whaley, 1998, Implied Volatility Smiles: Some Empirical Tests, {\it Journal of  Finance} {\bf 53}, 2059-2106.

\item [] Fama, E. F., 1963,  Mandelbrot and The Stable Paretian Hypothesis, {\it Journal of Business} {\bf 36}, 420-429.

\item [] \dash
%Fama, E. F.
, 1965, The Behaviour of Stock Market Prices, {\it Journal of Business} {\bf 38}, 34-105.

\item [] \dash
%Fama, E. F.
, 1991,  Efficient Capital Markets II, {\it Journal of Finance} {\bf 46}, 1575-1617.

\item [] Feller, W., 1954,  Diffusion processes in one dimension, {\it Transactions of the American Mathematical Society} {\bf 77}, 1-31.

\item [] Figlewski, S., 1989,  Options Arbitrage in Imperfect Markets, {\it Journal of  Finance} {\bf 64}, 1289-1311.

\item [] Gardiner, C. W., 1985, {\it Handbook of Stochastic Methods} (Springer-Verlag, Heilderberg, Berlin, New York).

\item [] Ghysels, E.,  A. C. Harvey, and E. Renault, 1996, Stochastic volatility, in G. S. Mandala and C. R. Rao eds: {\it Statistical Methods in Finance} (North-Holland, Amsterdam, New York).

\item [] Grossman, S. J., and J. E. Stiglitz, 1980, On the Impossibility of Informationally Efficient Markets, {\it The American Economic Review}  {\bf 70},  222-227.

%}

%\noindent
%\hspace{0.7 cm}
%\parbox{14.5cm}{

\item [] Harrison, J. M., and D. Kreps, 1979, Martingales and Multiperiod Securities Markets, {\it Journal of  Economic Theory} {\bf 20}, 381-408.

\item [] Harrison, J. M., and S. R. Pliska, 1981, Martingales and Stochastic Integrals in the Theory of Continuous Trading, {\it Stochastic Processes and Their Applications} {\bf 11}, 215-260.

\item [] Harrison, J. M., R. Pitbladdo and S. M. Schaefer, 1984, Continuous Price Processes in Frictionless Markets Have Infinite Variation, {\it Journal of Business} {\bf 57}, 353-365.

\item [] Heston, S., 1993, A Closed-Form Solution for Options with Stochastic Volatility to Bond and Currency Options, {\it Review of Financial Studies} {\bf 6}, 327-343.

\item [] Heston, S., and S. Nandi, 2000, A Closed-Form GARCH Option Valuation Model, {\it Review of Financial Studies} {\bf 13}, 585-625.

\item [] Hull, J. C., and A. White, 1987,  The Pricing of Options with Stochastic Volatilities, {\it Journal of  Finance} {\bf 42}, 281-300.

\item [] Hull, J. C., 2000, {\it Options, Futures, \& Other Derivatives} (Prentice-Hall, Upper Saddle River, NJ).

\item [] Lighthill, M. J., 1958, {\it An Introduction to Fourier Analysis and Generalized functions} (Cambridge University Press, Cambridge, UK).

\item [] Lo, A. W., and J. Wang, 1995, Implementing Option Pricing Models When Asset Returns Are Predictable, {\it Journal of Finance} {\bf 50}, 87-129.

\item [] Mandelbrot, B., 1963, The Variation of Certain Speculative Prices, {\it Journal of Business} {\bf 36}, 394-419.

\item [] Markowitz, H., 1952, Portfolio Selection, {\it Journal of  Finance} {\bf 7}, 77-91.

\item []  Merton, R. C., 1973a,  Theory of Rational Option Pricing, {\it Bell Journal of Economics} {\bf 4}, 141-183.

\item [] \dash
%Merton, R. C.
, 1973b, Appendix: Continuous-Time Speculative Processes, {\it SIAM Review} {\bf 1}, 34-38.

\item [] \dash
%Merton, R. C.
, 1976,  Option Pricing when Underlying Stock Returns are Discontinuous, {\it Journal of  Financial Economics} {\bf 3}, 125-144.

\item [] Mynt-U, Tyn, 1987, {\it Partial Differential Equations for Scientists and Engineers} (North-Holland, Amsterdam, New York).

\item [] Ross, S., 1976,  Arbitrage Theory of Capital Asset Pricing, {\it Journal of Economic Theory} {\bf 13}, 341-360.

\item []  Sharpe, W. F., 1964,  Capital Asset Prices, {\it Journal of Finance} {\bf 19}, 425-442.

\item [] Scott, L.O., 1987, Option Pricing when the Variance Changes Randomly: Theory, Estimation, and Application, {\it Journal of Financial and Quantitative Analysis} {\bf 22}, 419-438.

\item [] Stein, E. M., and J. C. Stein, 1991, Stock Price Distributions with Stochastic Volatility: An Analytic Approach, {\it Review of Financial Studies} {\bf 4}, 727-752.

\item [] Stratonovich, R. L., 1963, {\it Topics in the Theory of Random Noise} (Gordon and Breach, New York).

\item [] Uhlenbeck, G. E., and L. S. Ornstein, 1930,  On the Theory of the Brownian Motion, {\it Physical Review} {\bf 31}, 823-841.
%}
\endrefs

\end{document}